\documentclass[12pt]{article}

\usepackage[numbers, sort&compress]{natbib}
\usepackage{fullpage}
\usepackage{epsfig}
\usepackage{latexsym,amsmath,amssymb,amsthm}
\usepackage{subfig}
\newcommand{\ceil}[1]{\left\lceil {#1} \right\rceil}
\newcommand{\floor}[1]{\left\lfloor {#1} \right\rfloor}

\def\maps11{\stackrel {1-1}{\longmapsto}}

\begin{document}

\title{Scheduling strategies and throughput optimization for the Downlink for IEEE 802.11ax and IEEE 802.11ac based networks}

\author{%
Oran Sharon
\thanks{Corresponding author: oran@netanya.ac.il, Tel: 972-4-9831406,
Fax: 972-4-9930525} \\
Department of Computer Science \\
Netanya Academic College \\
1 University St. \\
Netanya, 42365 Israel
\and
Yaron Alpert\\
Intel\\
13 Zarchin St.\\
Ra'anana, 43662, Israel\\
Yaron.alpert@intel.com
% \and
% Chittabrata Ghosh\\
% Intel\\
% 2200 Mission College Blvd.\\
% Santa Clara, CA 95054, USA\\
% chittabrata.ghosh@intel.com\\
}

% \fi %%%%%

\date{}

\maketitle

\begin{abstract} 
The new IEEE 802.11 standard, IEEE 802.11ax, 
has the challenging
goal of serving more users compared
to its predecessor IEEE 802.11ac,
enabling consistent and reliable 
streams of data (average throughput) per station.
In this paper we explore some of the 
IEEE 802.11ax new mechanisms and
compare between the upper bounds on the throughputs 
of the Downlink unidirectional
UDP Multi Users (MU) triadic based 
on Multiple-Input-Multiple-Output (MU-MIMO) 
and Orthogonal Frequency Division Multiple Access (OFDMA)
transmission multiplexing format 
in IEEE 802.11ax vs. IEEE 802.11ac in 
the Single User (SU) and MU modes for 1, 4, 
8, 16, 32 and 64 stations scenario in reliable and unreliable
channels. The comparison is made
as a function of the Modulation and Coding
Schemes (MCS) in use. In IEEE 802.11ax 
we consider two flavors of
acknowledgment operation settings 
where the maximum acknowledgment
windows are 64 or 256 respectively. 
In SU scenario IEEE 802.11ax
upper bounds on the throughputs outperform IEEE 802.11ac 
by about 52$\%$ and 74$\%$ in reliable
and unreliable channels respectively. 
In MU-MIMO scenario
IEEE 802.11ax upper bounds on the throughputs outperform 
IEEE  802.11ac by about 59$\%$ and
103$\%$ in reliable and unreliable channels 
respectively.
Also, as the number
of stations increases, the advantage 
of IEEE 802.11ax in terms of the access
delay also increases.
\end{abstract}

\bigskip


\renewcommand{\baselinestretch}{1.3}
\small\normalsize

%%%%%%%%%%%%%%%%%%%%%%%%%%%%%%%%%%%%%%%%%%%%%%%%%%%%%%%%%%%%%%%%

\section{Introduction}

The latest IEEE 802.11 Standard (WiFi)~\cite{IEEEBase1}, 
created and maintained by 
the IEEE LAN/MAN Standards Committee (IEEE 802.11),
is currently the most effective solution within the range of Wireless Local
Area Networks (WLAN). Since its first release 
in 1997 the standard provides the basis 
for Wireless network products using 
the WiFi brand, and has since been 
improved upon in many ways. 
One of the main goals of these improvements
is to increase the throughput 
achieved by users and to improve
the standard's Quality-of-Service (QoS) capabilities. 
To fulfill the promise of increasing 
IEEE 802.11 performance and QoS capabilities,
a new amendment, IEEE 802.11ax ( also
known as High Efficiency (HE) ) was recently 
introduced~\cite{IEEEax}. IEEE 802.11ax is 
considered to be the sixth generation 
of a WLAN in the IEEE 802.11 set 
of types of WLANs and is 
a successor to IEEE 802.11ac~\cite{IEEEac,PS}.
The scope of the IEEE 802.11ax amendment is to
define modifications for both the 802.11 PHY and MAC
layers that enable at least four-fold improvement
in the average throughput per station in densely
deployed networks~\cite{KKL, AVA, DCC, B}.
Currently IEEE 802.11ax project is 
in a very early stage of development, 
due to be publicly released in 2019 .

In order to achieve its goals, one of 
the main challenges of IEEE 802.11ax is to enable simultaneous
transmissions by several stations and to enable
Quality-of-Service. Most of the research papers on IEEE 802.11ax
thus far deal with these challenges and examine different
access methods to enable efficient multi-user access
to random sets of stations.
For example, 
in~\cite{QLYY} the authors deal with the introduction
of Orthogonal Frequency Division Multiple Access (OFDMA)
into IEEE 802.11ax to enable multi user access.
They introduce an OFDMA based multiple access
protocol, denoted Orthogonal MAC for 802.11ax (OMAX),
to solve synchronization
problems and reduce overhead associated with
using OFDMA.
In~\cite{LLYQYZY} the authors suggest an access protocol
over the UL of an IEEE 802.11ax WLAN based on Multi User
Multiple-Input-Multiple-Output (MU-MIMO) and OFDMA PHY.
In~\cite{LDC} the authors suggest a centralized medium
access protocol for the UL of IEEE 802.11ax in order to efficiently
use the transmission resources.
In this protocol, stations transmit
requests for frequency sub-carriers, denoted
Resource Units (RU), to the AP over the UL. The AP
allocates RUs to the stations which use them later
for data transmissions over the UL.
In~\cite{KBPSL} a new method to use OFDMA over the UL
is suggested, where MAC Protocol Data Units (MPDU) from the stations are
of different lengths.
In~\cite{JS, RFBBO, RBFB, HYSG} a new version 
of the Carrier Sense Multiple Access with Collision Avoidance
(CSMA/CA) protocol, denoted Enhanced CSMA/CA (CSMA/ECA) is
suggested, which is suitable for IEEE 802.11ax . A deterministic
backoff is used after a successful transmission, and the backoff
stage is not reset after service. The backoff stage is reset
only when a station does not have any more MPDUs to transmit.
CSMA/ECA enables a more efficient use of the channel
and enhanced fairness.
In~\cite{KLL} the authors assume
a network with legacy and IEEE 802.11ax stations and examine
fairness issues between the two sets of the stations.

In this paper we do not suggest any new air access mechanisms
as the papers mentioned above do, but assume that the AP
is communicating in a regular fashion with a fixed set of stations.
The AP and the stations transmit in a Round Robin fashion,
without collisions.
We explore some of the Downlink (DL) and UL IEEE 802.11ax new 
mechanisms given that the AP knows with
which stations it communicates, and we compare 
between the upper bounds on the unidirectional 
UDP throughputs of IEEE 802.11ax and IEEE 802.11ac 
in Single User (SU) and Multi User (MU) modes
for 1, 4, 8, 16, 32 and 64 
stations scenarios in reliable and
unreliable channels. 
This is one of the aspects to compare between new
amendments of the IEEE 802.11 standard~\cite{KCC}.
We note that we do not assume that all the time
over the channel is devoted to UDP DL traffic.
It is possible that time is partitioned into intervals
of UDP DL traffic, UDP UL traffic, TCP traffic etc.
In this paper we investigate transmissions in the
time interval decvoted to UDP DL traffic.

In this paper we are interested in finding the upper bounds
on the throughputs
that can be achieved by IEEE 802.11ax and IEEE 802.11ac
and in comparing between the two. Therefore, we assume
the traffic saturation model
where all stations always have data to transmit.
Second, we neutralize any aspects
of the PHY layer as the relation between the Bit Error Rates (BER)
and the Modulation/Coding Scheme (MCS) in use, the number
of Spatial Streams (SS) in use, the channel correlation
when using MU-MIMO, i.e. we assume that there are
independent MU-MIMO channels for
each station, the use in sounding protocol etc.

The SU scenario 
implements sequential transmissions
in which a single wireless station sends
and receives data at every cycle one at 
a time, once it or the AP has gained access 
to the medium. The MU scenarios allow
for simultaneous transmission and reception to and 
from  multiple stations both in the 
DL and the UL directions. 
UL MU refers to simultaneous transmissions, i.e. at the same
time, from several stations to the AP over the UL.
The existing IEEE 802.11ac standard
does not enable UL MU while IEEE 802.11ax enables
up to 74 stations to transmit simultaneously over the UL.

The MU transmissions over the DL (DATA) and the UL (Acks) are done
by MIMO and OFDMA.
The IEEE 802.11ax standard
expends MIMO transmissions multiplexing format 
and specifies new ways of multiplexing additional users 
using OFDMA. 
The new IEEE 802.11ax OFDMA is backward compatible 
and enables scheduling different users in different
sub-carriers of the same channel. 
In the IEEE 802.11ac the total channel 
bandwidth (20 MHz, 40 MHz, 80 MHz etc. ) 
contains multiple OFDM sub-carriers. 
However, in IEEE 802.11ax OFDMA, different 
subsets of sub-carriers in the channel 
bandwidth can be used by different frame 
transmissions at the same time. Sub-carriers 
can be allocated for transmissions 
in Resource Units (RU)  as small as 2 MHz.

Given the above new structure of OFDMA in IEEE 802.11ax, 
the main contributions of this paper are as follows: First
we suggest several scheduling strategies by which the AP can
communicate with a set of stations over the DL. Second, we evaluate
upper bounds on
the throughput and the access delay performance of the
different scheduling strategies given the different PHY rates
of the RUs in the various scheduling strategies and
the different number of RUs in use, which influences the
PHY preamble's length.
This paper deals with the DL and a 
companion paper deals with the UL~\cite{SA2}.
The difference between the two papers
is in the direction in which
data is transmitted: in the current
paper the AP transmits data
to the stations, while in~\cite{SA2}
the stations transmit data to the AP.
As an outcome, the
current paper suggests scheduling
strategies for the transmission of data on the
DL, while~\cite{SA2} suggests scheduling
strategies for the transmission of data
on the UL. The strategies in the two papers
are different, using different features
of the IEEE 802.11ax amendment, e.g. different
control frames.

\indent
The remainder of the paper is organized as follows: 
In Section 2 we describe
the new mechanisms of IEEE 802.11ax 
relevant to this paper. In Section 3 we 
describe the transmission scenario 
by which we compare IEEE 802.11ax 
and IEEE 802.11ac in the SU and MU modes.
We assume that the reader is familiar 
with the basics of the PHY and MAC layers
of IEEE 802.11 described in previous papers, e.g.~\cite{SA}. 
In Section 4 we analytically compute 
the IEEE 802.11ax and IEEE 802.11ac 
throughputs. 
In Section 5 we make some approximations on the
amount of frame aggregation used in our transmission model.
In Section 6 we present the throughput 
of the various protocols and compare 
them. Section 7 summarizes 
the paper.
Lastly, 
we denote IEEE 802.11ac and IEEE 802.11ax 
by 11ac and 11ax respectively.

\section{The new features in IEEE 802.11ax}

IEEE 802.11ax focuses on implementing 
mechanisms to efficiently serve more
users, enabling consistent and 
reliable streams of data ( average throughput
per user ) in the presence of multiple users. 
Therefore, there are several 
new mechanisms in 11ax compared to 11ac both in the PHY and
MAC layers. At the PHY layer, 
11ax enables larger OFDM FFT sizes- 4X larger-
therefore every OFDM symbol is 
extended from $3.2 \mu s$ in 11ac to $12.8 \mu s$ in 11ax. 
By narrower subcarrier spacing (4X closer)
the protocol efficiency is increased, as
the same Guard Interval (GI) is used in both 11ax and 11ac .

To increase the average 
throughput per user in high-density scenarios,
11ax expands the 11ac Modulation Coding Schemes 
(MCSs) and adds MCS10 (1024 QAM ) and MCS 11 (1024 QAM 5/6), 
applicable for transmission with bandwidth larger than 20 MHz.

In this paper we focus on optimizing 
the IEEE 802.11 two-level aggregation
scheme working point first introduced 
in IEEE 802.11n~\cite{PS, IEEEBase1}, 
in which several  MPDUs can be aggregated 
to be transmitted in a  single PHY Service 
Data Unit (PSDU). Such aggregated PSDU 
is denoted Aggregate MAC Protocol Data
Unit (A-MPDU) frame. In two-level aggregation 
every MPDU can contain several 
MAC Service Data Units (MSDU). 
MPDUs are separated by an MPDU 
Delimiter field of 4 bytes and 
each MPDU contains MAC Header
and Frame Control Sequence (FCS) fields.
MSDUs within an MPDU are separated 
by a SubHeader field of 14 bytes. 
Every MSDU is rounded to an 
integral multiple of 4 bytes 
together with the SubHeader field. 
Every MPDU is also rounded to 
an integral multiple of 4 bytes.

\begin{figure}
\vskip 9cm
\includegraphics{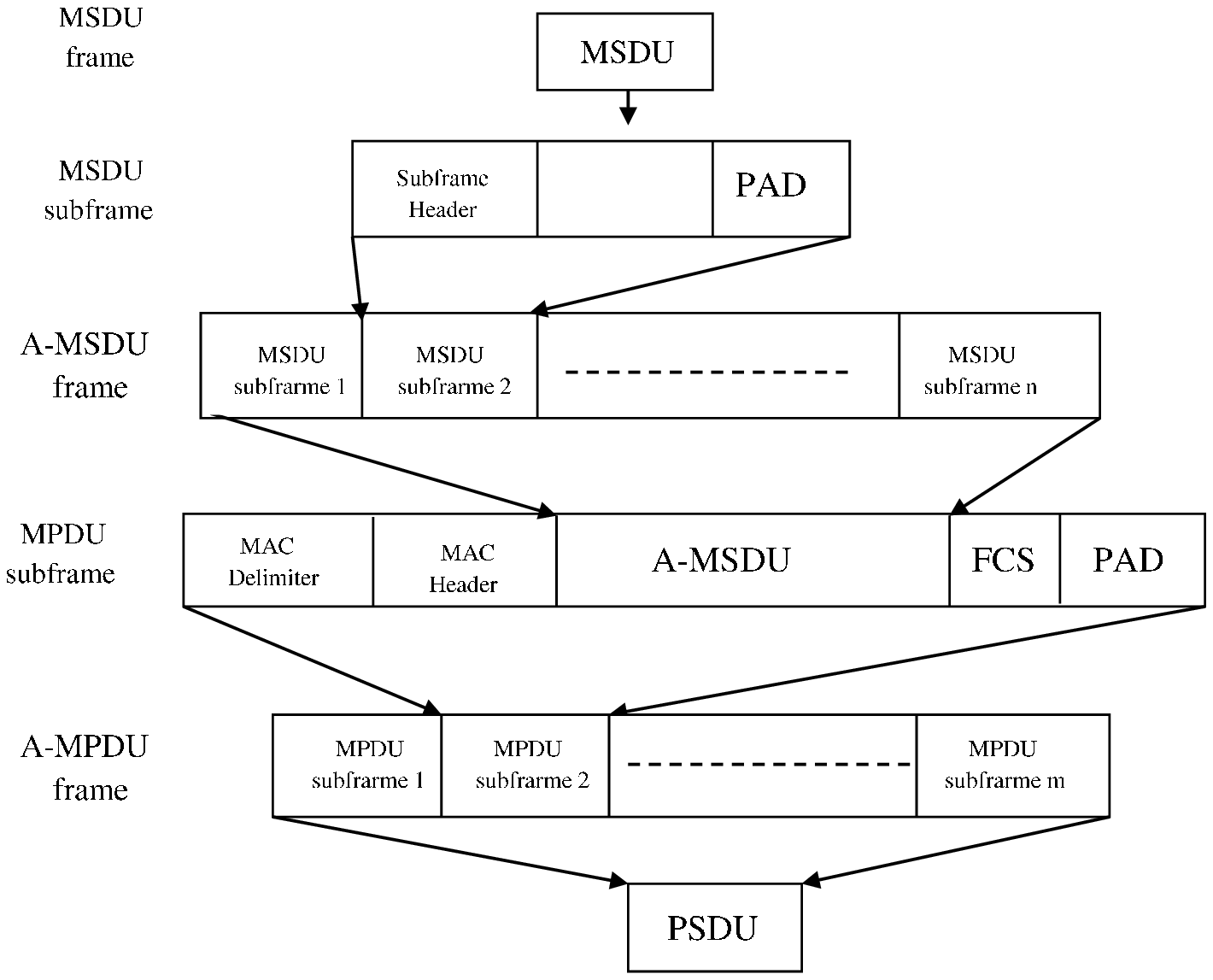}
\caption{The generation of an A-MPDU frame
in two-level aggregation.}
\label{fig:twole}
\end{figure}

In 11ax and 11ac the size of 
an MPDU is limited to 11454 bytes. In 11ac an A-MPDU
is limited to 1,048,575 bytes and 
this limit is extended to 4,194,304 bytes 
in 11ax. In both 11ac and 11ax 
the transmission time of the PPDU 
(PSDU and its preamble) is limited to 
$5.484ms$ ($5484 \mu s$)
due to the L-SIG (one of the legacy 
preamble's fields) duration limit~\cite{IEEEBase1}.
The A-MPDU frame structure in two-level aggregation
is shown in Figure~\ref{fig:twole}.

11ax also enables extension of the acknowledgment 
mechanism by using a 256 maximum 
acknowledgment window vs. maximum 
window of 64 in 11ac. In this paper 
we also assume that all MPDUs 
transmitted in an A-MPDU frame are 
from the same Traffic Stream (TS). 
In this case up to 256 MPDUs are
allowed in an A-MPDU frame of 11ax, 
while in 11ac up to only 64 MPDUs
are allowed. 

Finally, in 11ac it 
is possible to transmit simultaneously 
up to 4 stations only over the DL 
using MU. In 11ax this number 
is extended to 74. Also, in 11ax 
it is possible to transmit by 
MU-MIMO or OFDMA 
both over DL and
UL, while in 11ac only UL SU mode is supported.

\section{Model}

\subsection{Transmission patterns}

\begin{figure}
\vskip 14cm
\includegraphics{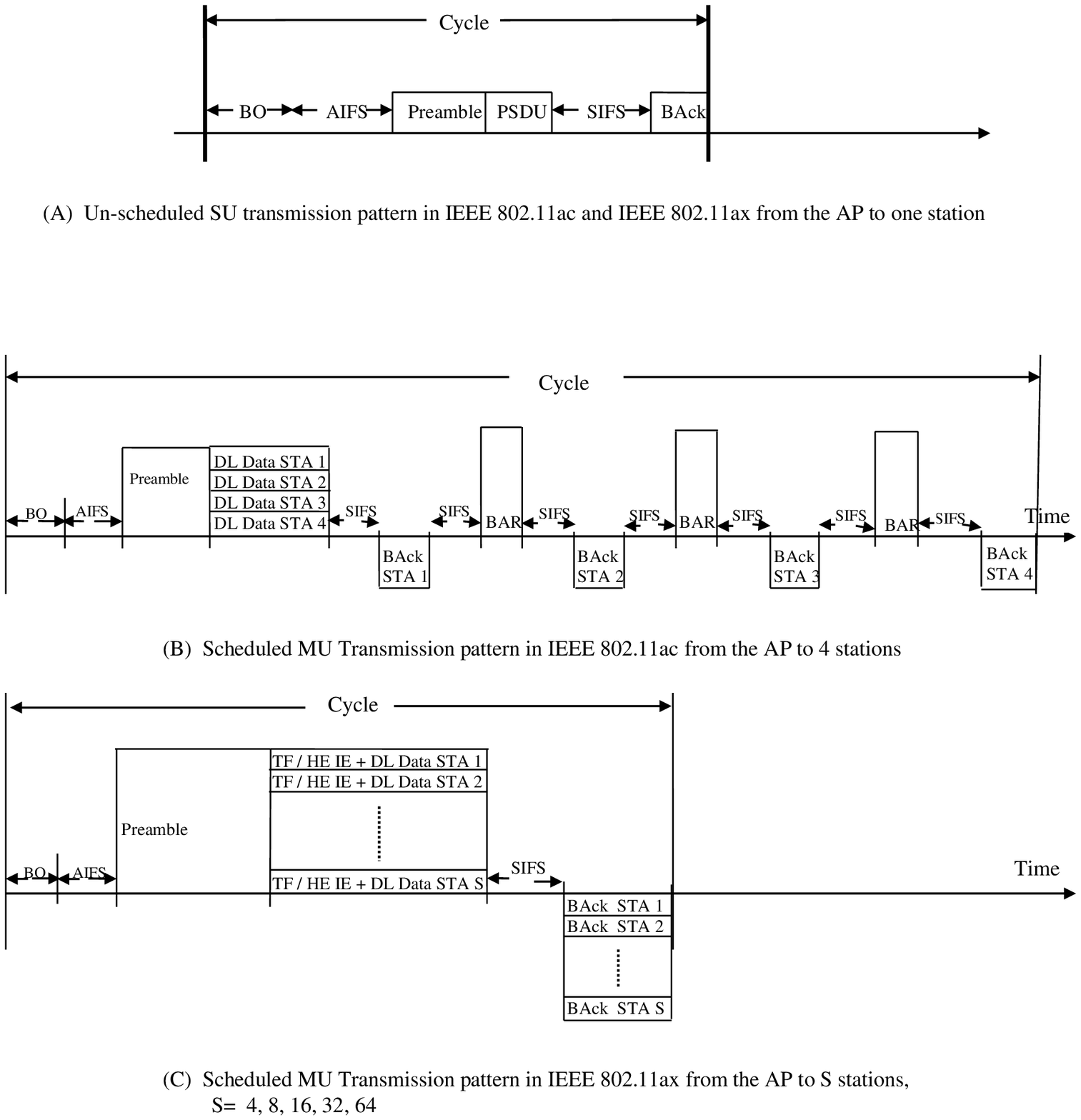}
\caption{Transmissions from the AP to stations in Single User and Multi User modes in IEEE 802.11ac and in IEEE 802.11ax .}
\label{fig:traffic}
\end{figure}

As mentioned, one of the main goals of 11ax is to enable
larger throughputs in the network when transmitting
to several stations. In 11ax it is possible to transmit/receive
simultaneously to/from 74 stations over the DL/UL
while in 11ac the number
of stations is limited to 4, and only over the
DL. In this paper we compare
the throughputs received in 11ac and 11ax
when transmitting to $S$ stations, $S=1, 4, 8, 16, 32$ and
64 stations. Transmitting to one station only
is done by using the SU mode
of transmissions. The AP transmits to one station
and receives a Block Ack (BAck) frame in return. In this mode
the advantage of 11ax over 11ac is in its more
efficient PHY layer and its new MCSs. The unscheduled SU traffic
pattern in this case is shown in Figure~\ref{fig:traffic}(A)
for both 11ac and 11ax.

\begin{figure}
\vskip 10cm
\includegraphics{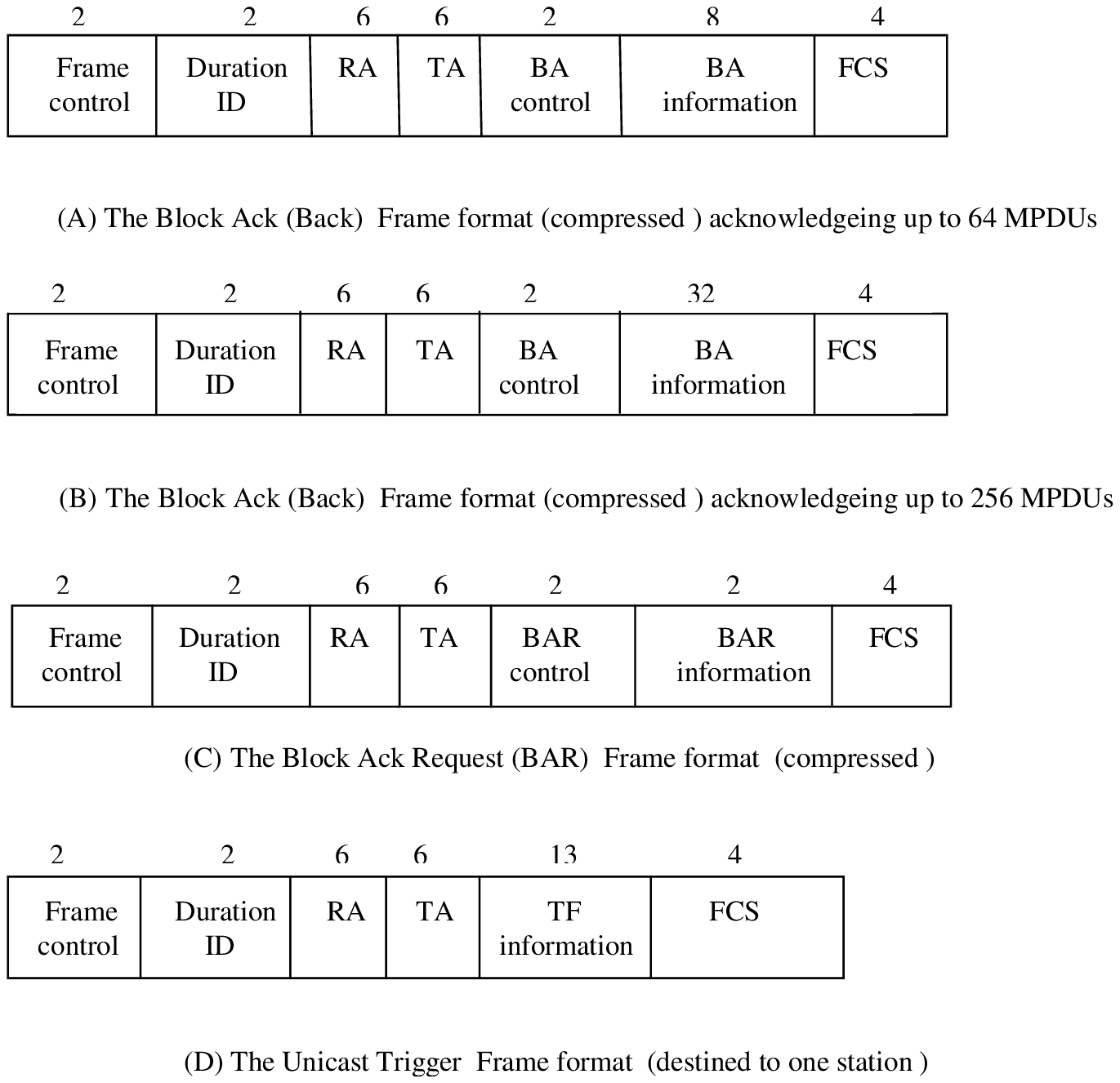}
\caption{The Block Ack (BAck), the Block Ack request (BAR)
and the Trigger Frame (TF) frames' format.}
\label{fig:frameformat}
\end{figure}

Transmitting to several stations can be done
in two ways. The first is by SU mode.
When transmitting to $S$ stations, the transmission cycle
in Figure~\ref{fig:traffic}(A) repeats itself $S$ times.
Another alternative is to use MU mode
in which the AP transmits simultaneously to several stations
in the same transmission opportunity over the channel.
In Figure~\ref{fig:traffic}(B) we show this possibility for 11ac
where the AP transmits to 4 stations simultaneously. 
This is the maximum
number of stations to which the AP can transmit
simultaneously in 11ac. 
In UL the stations transmit 4 sequential
BAck frames using the Single User (SU) legacy
mode. While the first BAck is transmitted
SIFS immediately after receiving
the transmission from the AP,
the last 3 are solicited
by BAck Request (BAR) frames from the AP.
Each BAR is transmitted
SIFS after the previous BAck. 
The formats of the BAck and BAR frames are shown
in Figures~\ref{fig:frameformat}(A),~\ref{fig:frameformat}(B) 
and~\ref{fig:frameformat}(C) respectively.

% \begin{figure}
% \vskip 5cm
% \special{psfile=frameformat2.ps voffset =-370 hoffset= -10 hscale = 80 vscale = 80}
% \caption{The Block Ack Request (BAR) and Trigger Frame (TF) frame formats.}
% \label{fig:frameformat2}
% \end{figure}

In 11ax, Figure~\ref{fig:traffic}(C),
the AP transmits over
the DL to $S$ stations simultaneously using
MU-MIMO or OFDMA or combination, as in 11ac, 
and the stations transmit their BAck frames
simultaneously in the UL using MU-MIMO or OFDMA or
a combination. This
is possible only in 11ax .
The AP allocates the UL Resource Units (RU), i.e. subchannels in the
case of OFDMA and Frequency/Spatial
Streams in the case of MU-MIMO,
for the transmissions of the stations,
by one of two possible UL RU allocation signaling methods:
In the first method
the AP transmits a
unicast Trigger
Frame (TF) to every station that contains the UL RU allocation. 
This frame is a control MAC Protocol Data Unit
(MPDU) that is added to the other Data MPDUs 
which the AP transmits to a station in an A-MPDU
frame.  The format of the TF
frame is shown in Figure~\ref{fig:frameformat}(D).
For a unicast TF the TF information field contains two sub-fields:
one is a common part of 8 bytes and the second is a user element 
of 4 bytes.
The other alternative
method is
to add an HE Control Element to $every$
Data MPDU in the A-MPDU frame that is transmitted to every station.
In the following throughput computations we optimize
the amount of overhead used due to the above methods
by computing the minimum overhead needed as a function
of the number of data MPDUs in the A-MPDU frame.

Finally, we assume that the AP and the stations do not
contend for the channel and so there are no collisions. The
cycles in Figure~\ref{fig:traffic}(A), (B) and (C) repeat one after
the other. This is possible by e.g. configuring the stations
in a way that prevents collisions. For example, the stations are
configured to choose their BackOff intervals from very large
contention interval, other than the defaults ones~\cite{IEEEBase1}. 
Thus, the AP always wins the channel without collisions.

\subsection{DL service transmissions' scheduling strategies}

There are several DL service scheduling strategies to transmit
to a group of stations, and we compare between
them. We now specify these scheduling strategies for every
number $S$ of stations, $S=1, 4, 8, 16, 32, 64$.
By $x \cdot SU_{AX}(1)$ and $x \cdot SU_{AC}(1)$
we denote a transmission to $n$ stations
in 11ax and 11ac respectively,
using the transmission pattern in Figure~\ref{fig:traffic}(A)
$x$ times in sequence, every transmission is to a different
station. By $x \cdot MU_{AC}(4)$ we denote transmissions to $4x$
stations using the traffic pattern of Figure~\ref{fig:traffic}(B)
$x$ times in sequence, every transmission is to a different
group of 4 stations. By $m \cdot MU_{AX}(n)$ we denote transmissions
to $m \cdot n$ stations using the traffic pattern 
of Figure~\ref{fig:traffic}(C) $m$ times in sequence, each transmission
to a different group of $n$ stations. In this paper
$n=4, 8, 16, 32$ and 64.

\noindent
The DL service scheduling strategies are as follows:

\begin{itemize}

\item
$S=1$:

\noindent
11ac : $1 \cdot SU_{AC}(1)$ .

\noindent
11ax : $1 \cdot SU_{AX}(1)$ .

\item
$S=4$:

\noindent
11ac : $4 \cdot SU_{AC}(1)$, $1 \cdot MU_{AC}(4)$.

\noindent
11ax : $4 \cdot SU_{AX}(1)$, $1 \cdot MU_{AX}(4)$.

\item
$S=8$:

\noindent
11ac : $8 \cdot SU_{AC}(1)$, $2 \cdot MU_{AC}(4)$.

\noindent
11ax : $8 \cdot SU_{AX}(1)$, $2 \cdot MU_{AX}(4)$, $1 \cdot MU_{AX}(8)$ .

\item
$S=16$:

\noindent
11ac : $16 \cdot SU_{AC}(1)$, $4 \cdot MU_{AC}(4)$ .

\noindent
11ax : $16 \cdot SU_{AX}(1)$, $4 \cdot MU_{AX}(4)$, $2 \cdot MU_{AX}(8)$, $1 \cdot MU_{AX}(16)$.

\item
$S=32$:

\noindent
11ac : $32 \cdot SU_{AC}(1)$, $8 \cdot MU_{AC}(4)$ .

\noindent
11ax : $32 \cdot SU_{AX}(1)$, $8 \cdot MU_{AX}(4)$, $4 \cdot MU_{AX}(8)$, $2 \cdot MU_{AX}(16)$, $1 \cdot MU_{AX}(32)$.

\item
$S=64$:

\noindent
11ac : $64 \cdot SU_{AC}(1)$, $16 \cdot MU_{AC}(4)$ .

\noindent
11ax : $64 \cdot SU_{AX}(1)$, $16 \cdot MU_{AX}(4)$, $8 \cdot MU_{AX}(8)$, $4 \cdot MU_{AX}(16)$, $2 \cdot MU_{AX}(32)$, $1 \cdot MU_{AX}(64)$.

\end{itemize}

\subsection{Channel assignment}

We assume
the 5GHz band, a 160MHz channel, the AP has 4 antennas and every station
has 1 antenna.
In SU(1) and in the DL direction
the entire channel is devoted to transmissions of the AP
in both 11ac and 11ax .
In UL 
SU the BAck frame is transmitted by using the legacy
PHY basic rates. Therefore the UL Ack is sent at
legacy mode where the station
is transmitting in a 20 MHz primary channel and its transmission
is duplicated 8 times in order to occupy the entire 160 MHz.
The UL PHY rate is set to the
largest possible PHY rate in the set that
is smaller or equal to the DL Data rate.

When using MU mode the 160MHz channel
is divided into $\frac{S}{4}$ channels of
$\frac{160 \cdot 4}{S}$ MHz each, $S=4, 8, 16, 32, 64$. The
AP transmits to 4 stations in every such channel,
using 4 Spatial Streams. For example, for $S=64$
there are 16 channels of 10MHz each; in each of them
the AP transmits to 4 stations.
When $S=4$ only MU-MIMO is used. For $S>4$ MU-MIMO+OFDMA is used.
In the case of $MU_{AC}$, Figure~\ref{fig:traffic}(B),
it is again possible to transmit 
the Back frames 
in the UL direction 
only in the legacy mode, as in SU(1),
and the UL PHY rate is set again to
the largest possible
PHY basic rate in the set that is smaller or equal
to the DL Data rate. 
Again, the primary 20 MHz channel
is duplicated 8 times in all
secondary channels to occupy the entire 160 MHz channel.

For the UL Ack transmission in 11ax, Figure~\ref{fig:traffic}(C),
we assume either MU-MIMO or OFDMA.
In the case of UL MU-MIMO the transmissions
are symmetrical to those in DL.
In the case of UL OFDMA the 160 MHz channel
is divided into $S$ channels of $\frac{160}{S}$MHz each,
except in the case of $S=64$ where each station
is allocated a channel of 2 MHz. 

\subsection{PPDU formats}

\begin{figure}
\vskip 12cm
\includegraphics{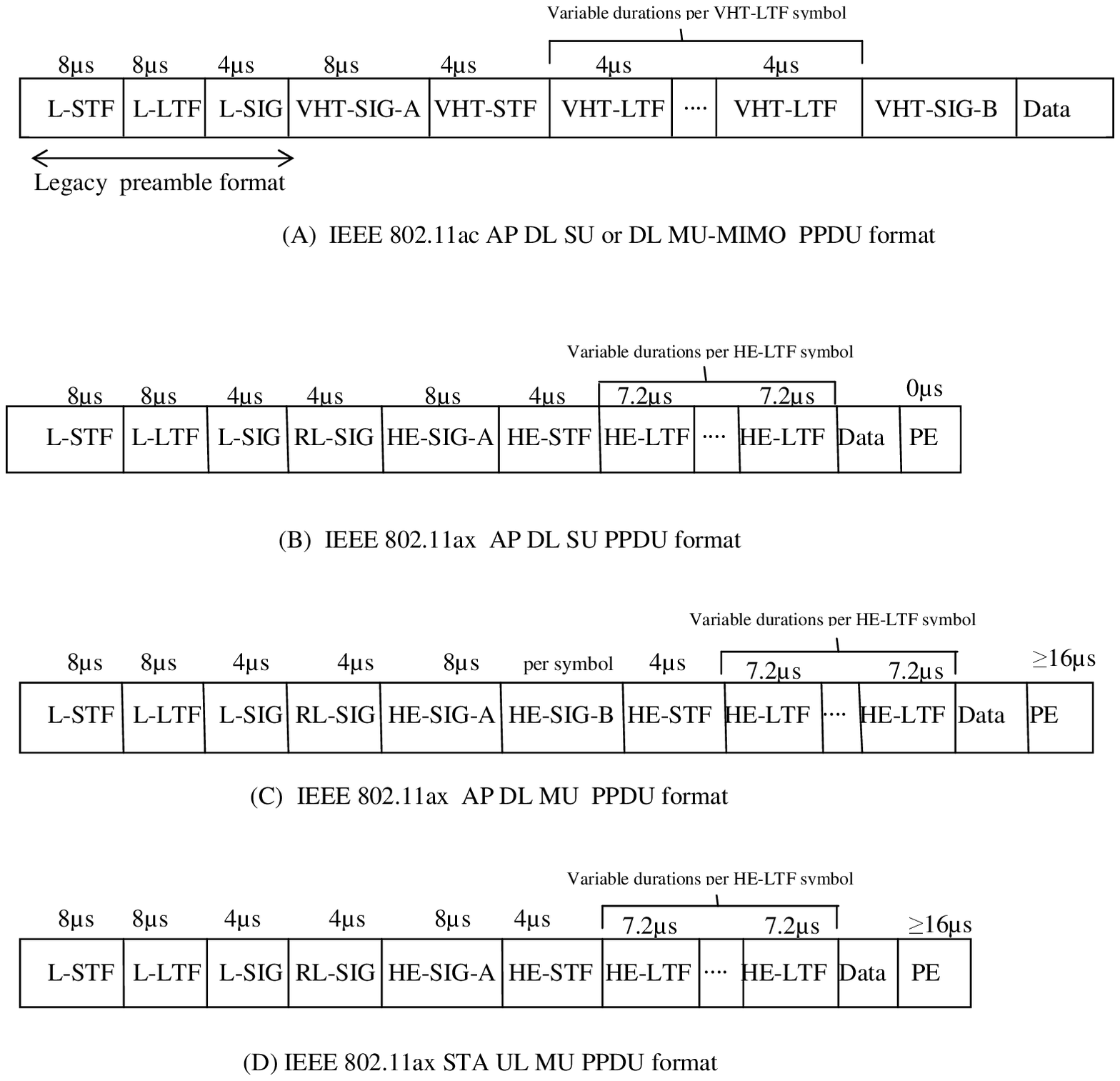}
\caption{The PPDU formats in the SU and MU modes.}
\label{fig:formatPPDU}
\end{figure}

In Figure~\ref{fig:formatPPDU} we show the various
PPDUs' formats in use in the
various transmission patterns of Figure~\ref{fig:traffic}.

In Figures~\ref{fig:formatPPDU}(A) and~\ref{fig:formatPPDU}(B) 
we show the PPDU formats
used in the DL SU of 11ac and 11ax respectively,
Figure~\ref{fig:traffic}(A).
In the PPDU format of 11ac are
the VHT-LTF fields, the number of which equals
the number of SS in use and each is $4 \mu s$.
In the 11ax PPDU format there are the HE-LTF fields,
the number of which equals again to the number of SS
in use. In this paper we assume that each such field is
composed of 2X LTF and therefore of duration
$7.2 \mu s$~\cite{IEEEax}.
Notice that in SU mode and when using the
same number $X$ of SS, the
preamble in 11ax is longer than that
in 11ac by $4 \mu s + X \cdot (7.2-4) \mu s= 4 \mu s + X \cdot 3.2 \mu s$.

Notice also that the PSDU frame in 11ax contains
a Packet Extension (PE) field.
This field is mainly used in Multi-User (MU) mode
and we assume it is not present in SU, i.e. 
it is of length $0 \mu s$.

In Figures~\ref{fig:formatPPDU}(A) and~\ref{fig:formatPPDU}(B) we also
show the legacy preamble, used in both 11ac and 11ax in the
UL SU. 

The PPDU format in Figure~\ref{fig:formatPPDU}(A) is
also used in the DL MU-MIMO in 11ac.
In Figure~\ref{fig:formatPPDU}(C)
we show the PPDU format used in 11ax in DL MU.
In this frame format
there are again the HE-LTF fields, the
number of which equals the number of SS.
As in the SU mode we assume each such field is composed
of 2X LTF
and therefore is of 
duration $7.2 \mu s$.
The MCS used 
in the HE-SIG-B field
is the
minimum between MCS4 and the one used for the data
transmissions~\cite{IEEEax}. The length of this
field is also a function of the number
of stations to which the AP transmits
simultaneously. Therefore, in the case of e.g. 4 stations
the HE-SIG-B field duration is $8 \mu s$ for MCS0 and MCS1,
and is $4 \mu s$ for MCS2-4
following section 29.3.9.8 in~\cite{IEEEax}.
For MCS5-MCS11 it is $4 \mu s$ as for MCS4.

In Figure~\ref{fig:formatPPDU}(D) we
show the PPDU format used in UL MU
in 11ax which is used in the traffic pattern of Figure~\ref{fig:traffic}(C).
Notice again that in 11ax the PSDU is followed
by a Packet Extension (PE) field which is used
to enable the receiver of the PSDU additional time
to move from a reception mode to a transmission mode.
The largest duration of this field is $16 \mu s$ which
we assume in this paper.

\subsection{Parameters' values}

In Table~\ref{tab:phyratesSU} we show
the PHY rates and the preambles
used in 11ac and 11ax in SU mode and
in the various MCSs. 
In Table~\ref{tab:phyrates} we show the PHY rates and the preambles
used in 11ac and 11ax in MU mode, in the various MCSs and
in all cases of the number of stations $S$, i.e. 
$S= 4, 8, 16, 32$ and 64.
The values in both tables are taken from~\cite{IEEEax}.

\begin{table}
\caption{\label{tab:phyratesSU}{The PHY rates and the preambles in the DL and UL
of IEEE 802.11ac and IEEE 802.11ax in the case of a 160 MHz channel, 1 Spatial
Stream and legacy UL channel. Single User mode.}}
\vspace{3 mm}
\tiny
\center
\begin{tabular}{|r|c|c|c|c|c|c|c|c|c|}  \hline
     & \multicolumn{2}{|c|}1 &  \multicolumn{2}{|c|}2 & &  
       \multicolumn{2}{|c|}3 & \multicolumn{2}{|c|}4 \\ \cline{2-10}
     & \multicolumn{2}{c|}{SU DL data}&  \multicolumn{2}{c|}{SU DL data}&  &
       \multicolumn{2}{c|}{UL BAck} & \multicolumn{2}{c|}{UL BAck} \\
     & \multicolumn{2}{c|}{transmission rate in 11ax}  &  \multicolumn{2}{c|}{transmission rate in 11ac}& & \multicolumn{2}{c|}{transmission rate in 11ax} & \multicolumn{2}{c|}{transmission rate in 11ac}  \\ \hline
     &  PHY Rate & Preamble & 
       PHY Rate & Preamble & & PHY Rate & Preamble & PHY rate & Preamble \\
MCS  &  (Mbps)   & ($\mu s$)   
     &  (Mbps)    & ($\mu s$) & &  (Mbps)   & ($\mu s$) & (Mbps) & ($\mu s$) \\
     &  GI$=0.8 \mu s$   &
     &   GI$=0.8 \mu s$   &   & &  & & & \\ \hline
     & \multicolumn{2}{c|}{1 station IEEE 802.11 ax} &\multicolumn{2}{c|}{ 1 station IEEE 802.11 ac} & & & & &  \\ \hline
 0   &   72.1 &  43.2 &   58.5 & 36.0 & &48.0 & 20.0 & 48.0 & 20.0\\ 
 1   &  144.1 &  43.2 &  117.0 & 36.0 & &48.0 & 20.0 & 48.0 & 20.0\\ 
 2   &  216.2 &  43.2 &  175.5 & 36.0 & &48.0 & 20.0 & 48.0 & 20.0\\ 
 3   &  288.2 &  43.2 &  234.0 & 36.0 & &48.0 & 20.0 & 48.0 & 20.0\\ 
 4   &  432.4 &  43.2 &  351.0 & 36.0 & &48.0 & 20.0 & 48.0 & 20.0\\ 
 5   &  576.5 &  43.2 &  468.0 & 36.0 & &48.0 & 20.0 & 48.0 & 20.0\\ 
 6   &  648.5 &  43.2 &  526.5 & 36.0 & &48.0 & 20.0 & 48.0 & 20.0\\ 
 7   &  720.6 &  43.2 &  585.0 & 36.0 & &48.0 & 20.0 & 48.0 & 20.0\\ 
8    &  864.7 &  43.2 &  702.0 & 36.0 & &48.0 & 20.0 & 48.0 & 20.0\\ 
 9   &  960.7 &  43.2 &  780.0 & 36.0 & &48.0 & 20.0 & 48.0 & 20.0\\ 
10   & 1080.9 &  43.2 &  N/A   & N/A & &48.0 & 20.0 & N/A & N/A \\ 
11   & 1201.0 &  43.2 &  N/A   & N/A & &48.0 & 20.0 & N/A & N/A \\ \hline 
\end{tabular}  
\end{table}

\begin{table}
\caption{\label{tab:phyrates}{The PHY rates and the preambles in the DL and UL
of IEEE 802.11ac and IEEE 802.11ax in the case of a 160 MHz channel, 4 Spatial
Streams and legacy UL channel in IEEE 802.11ac . Multi User mode.}}
\vspace{3 mm}
\tiny
\center
\begin{tabular}{|r|c|c|c|c|c|c|c|c|c|c|c|}  \hline
     & \multicolumn{2}{|c|}1 &  \multicolumn{2}{|c|}2  &
       \multicolumn{2}{|c|}3 & & \multicolumn{2}{|c|}4  &  
       \multicolumn{2}{|c|}5  \\ \cline{2-12}
     & \multicolumn{2}{c|}{DL MU data}&  \multicolumn{2}{c|}{UL MU-MIMO BAck} & \multicolumn{2}{c|}{UL OFDMA BAck}& & \multicolumn{2}{c|}{DL MU-MIMO data}&  \multicolumn{2}{c|}{UL BAck} \\
     & \multicolumn{2}{c|}{transmission rate in 11ax}&  \multicolumn{2}{c|}{transmission rate in 11ax}& \multicolumn{2}{c|}{transmission rate in 11ax}  & & \multicolumn{2}{c|}{transmission rate in 11ac}&  \multicolumn{2}{c|}{transmission rate in 11ac}  \\ \hline
     &  PHY Rate &  Preamble & PHY Rate & Preamble & PHY Rate & Preamble &
     &  PHY Rate & Preamble &  PHY Rate & Preamble \\
MCS  & (MBps)    &  ($\mu s$)  &  (MBps)  & ($\mu s$)  & (MBps)   & ($\mu s$) 
     & & (MBps)    & ($\mu s$)  &  (MBps)   & ($\mu s$) \\
     &  GI$=0.8 \mu s$   &   & GI$=1.6 \mu s$  &  & GI$=1.6 \mu s$  & 
     &  & GI$=0.8 \mu s$   &    &  & \\ \hline
     & \multicolumn{6}{c|}{4 stations IEEE 802.11 ax} &  &\multicolumn{4}{c|}{ 4 stations IEEE 802.11 ac} \\ \hline
 0   & 72.1   & 72.8 & 68.1   & 64.8  & 16.3  &  64.8 & &58.5   & 48.0 & 48.0 &  20.0\\ 
 1   & 144.1  & 72.8 & 136.1  & 64.8  & 32.5  &  64.8 & &117.0  & 48.0 & 48.0 &  20.0\\ 
 2   & 216.2  & 68.8 & 204.2  & 64.8  & 48.8  &  64.8 & &175.5  & 48.0 & 48.0 & 20.0 \\ 
 3   & 288.2  & 68.8 & 272.2  & 64.8  & 65.0  &  64.8 & &234.0  & 48.0 & 48.0 & 20.0 \\ 
 4   & 432.4  & 68.8 & 408.3  & 64.8  & 97.5  &  64.8 & &351.0  & 48.0 & 48.0 &  20.0\\ 
 5   & 576.5  & 68.8 & 544.4  & 64.8  & 130.0 &  64.8 & &468.0  & 48.0 & 48.0 & 20.0 \\ 
 6   & 648.5  & 68.8 & 612.5  & 64.8  & 146.3 &  64.8 & &526.5  & 48.0 & 48.0 & 20.0 \\ 
 7   & 720.6  & 68.8 & 680.6  & 64.8  & 162.5 &  64.8 & &585.0  & 48.0 & 48.0 & 20.0 \\ 
8    & 864.7  & 68.8 & 816.7  & 64.8  & 195.0 &  64.8 & &702.0  & 48.0 & 48.0 & 20.0 \\ 
 9   & 960.7  & 68.8 & 907.4  & 64.8  & 216.7 &  64.8 & &780.0  & 48.0 & 48.0 & 20.0 \\ 
10   & 1080.9 & 68.8 & 1020.8 & 64.8  & 243.8 &  64.8 & & N/A   & N/A & N/A & N/A \\ 
11   & 1201.0 & 68.8 & 1134.2 & 64.8  & 270.8 &  64.8 & & N/A   & N/A & N/A & N/A \\ \hline 
     & \multicolumn{6}{c|}{8 stations IEEE 802.11 ax} &  &\multicolumn{4}{c|}{ 4 stations IEEE 802.11 ac} \\ \hline
 0   & 36.0   & 76.8 & 34.0   & 64.8  & 8.1  &  64.8 & &58.5   & 48.0  & 48.0 &  20.0 \\ 
 1   & 72.1  & 76.8 & 68.1  & 64.8  & 16.3  &  64.8 & &117.0  & 48.0   & 48.0 & 20.0 \\ 
 2   & 108.1  & 72.8 & 102.1  & 64.8  & 24.4  &  64.8 & &175.5  & 48.0 & 48.0 & 20.0 \\ 
 3   & 144.1  & 72.8 & 136.1  & 64.8  & 32.5  &  64.8 & &234.0 & 48.0  & 48.0 & 20.0 \\ 
4   & 216.2  & 68.8 & 204.2  & 64.8  & 48.8  &  64.8 & &351.0 & 48.0   & 48.0 &  20.0\\ 
 5   & 288.2  & 68.8 & 272.2  & 64.8  & 65.0 &  64.8 & &468.0  & 48.0  & 48.0 &  20.0\\ 
 6   & 324.3  & 68.8 & 306.3  & 64.8  & 73.1 &  64.8 & &526.5  & 48.0  & 48.0 &  20.0\\ 
 7   & 360.3  & 68.8 & 340.3  & 64.8  & 81.3 &  64.8 & &585.0  & 48.0  & 48.0 &  20.0\\ 
8    & 432.4  & 68.8 & 408.3  & 64.8  & 97.5 &  64.8 & &702.0  & 48.0  & 48.0 & 20.0  \\ 
 9   & 480.4  & 68.8 & 453.7  & 64.8  & 108.3 &  64.8 & &780.0  & 48.0 &48.0 & 20.0 \\ 
10   & 540.4 & 68.8 & 510.4 & 64.8  & 121.9 &  64.8 &  &N/A   & N/A & N/A & N/A \\ 
11   & 600.4 & 68.8 & 567.1 & 64.8  & 135.4 &  64.8 &  &N/A   & N/A & N/A & N/A \\ \hline 
     & \multicolumn{6}{c|}{16 stations IEEE 802.11 ax} &  &\multicolumn{4}{c|}{ 4 stations IEEE 802.11 ac} \\ \hline
 0   & 17.2   & 84.8 & 16.3   & 64.8  & 8.1  &  64.8 & &58.5   & 48.0 & 48.0 &  20.0 \\ 
 1   & 34.4  & 84.8 & 32.5  & 64.8  & 16.3  &  64.8 & &117.0  & 48.0 & 48.0 &   20.0 \\ 
 2   & 51.6  & 76.8 & 48.8  & 64.8  & 24.4  &  64.8 & &175.5  & 48.0 & 48.0 &   20.0 \\ 
 3   & 68.8  & 76.8 & 65.0  & 64.8  & 32.5  &  64.8 & &234.0  & 48.0 & 48.0 &   20.0       \\ 
 4   & 103.2  & 72.8 & 97.5  & 64.8  & 48.8  &  64.8 & &351.0  & 48.0 & 48.0 &   20.0 \\ 
 5   & 137.6  & 72.8 & 130.0  & 64.8  & 65.0 &  64.8 & &468.0  & 48.0 & 48.0 &   20.0  \\ 
 6   & 154.9  & 72.8 & 146.3  & 64.8  & 73.1 &  64.8 & &526.5  & 48.0 & 48.0 &  20.0  \\ 
 7   & 172.1  & 72.8 & 162.5  & 64.8  & 81.3 &  64.8 & &585.0  & 48.0 & 48.0 &  20.0  \\ 
8    & 206.5  & 72.8 & 195.0  & 64.8  & 97.5 &  64.8 & &702.0  & 48.0 & 48.0 &  20.0  \\ 
 9   & 229.4  & 72.8 & 216.7  & 64.8  & 108.3 &  64.8 & &780.0  & 48.0 & 48.0 &  20.0 \\ 
10   & 258.1 & 72.8 & 243.8 & 64.8  & N/A & N/A  &  &N/A   & N/A & N/A & N/A  \\ 
11   & 286.8 & 72.8 & 270.8 & 64.8  & N/A & N/A  &  &N/A   & N/A & N/A & N/A  \\ \hline 
\end{tabular}  
\end{table}

\addtocounter{table}{-1}

\begin{table}
\caption{\label{tab:phyrates1}{(cont.)}}
\vspace{3 mm}
\tiny
\center
\begin{tabular}{|r|c|c|c|c|c|c|c|c|c|c|c|}  \hline
     & \multicolumn{2}{|c|}1 &  \multicolumn{2}{|c|}2  &
       \multicolumn{2}{|c|}3 & & \multicolumn{2}{|c|}4  &  
       \multicolumn{2}{|c|}5  \\ \cline{2-12}
     & \multicolumn{2}{c|}{DL MU data}&  \multicolumn{2}{c|}{UL MU-MIMO BAck} & \multicolumn{2}{c|}{UL OFDMA BAck}& & \multicolumn{2}{c|}{DL MU-MIMO data}&  \multicolumn{2}{c|}{UL BAck} \\
     & \multicolumn{2}{c|}{transmission rate in 11ax}&  \multicolumn{2}{c|}{transmission rate in 11ax}& \multicolumn{2}{c|}{transmission rate in 11ax}  & & \multicolumn{2}{c|}{transmission rate in 11ac}&  \multicolumn{2}{c|}{transmission rate in 11ac}  \\ \hline
     &  PHY Rate &  Preamble & PHY Rate & Preamble & PHY Rate & Preamble &
     &  PHY Rate & Preamble &  PHY Rate & Preamble \\
MCS  & (MBps)    &  ($\mu s$)  &  (MBps)  & ($\mu s$)  & (MBps)   & ($\mu s$) 
     & & (MBps)    & ($\mu s$)  &  (MBps)   & ($\mu s$) \\
     &  GI$=0.8 \mu s$   &   & GI$=1.6 \mu s$  &  & GI$=1.6 \mu s$  & 
     &  & GI$=0.8 \mu s$   &    &  & \\ \hline
     & \multicolumn{6}{c|}{32 stations IEEE 802.11 ax} &  &\multicolumn{4}{c|}{ 4 stations IEEE 802.11 ac} \\ \hline
 0   & 8.6  & 104.8 & 8.1 & 64.8  & 1.7 &  64.8 & &58.5   & 48.0 & 48.0 &  20.0\\ 

 1   & 17.2 & 104.8 & 16.3 & 64.8  & 3.3  &  64.8 & &117.0  & 48.0 & 48.0 &  20.0  \\ 
 2   & 25.8  & 84.8 & 24.4 & 64.8  & 5.0  &  64.8 & &175.5  & 48.0 & 48.0 &  20.0  \\ 
 3   & 34.4  & 84.8 & 32.5 & 64.8  & 6.7  &  64.8 & &234.0  & 48.0 & 48.0 &  20.0  \\ 
 4   & 51.6  & 80.8 & 48.8 & 64.8  & 10.0 &  64.8 & &351.0  & 48.0 & 48.0 &  20.0  \\ 
 5   & 68.8  & 80.8 & 65.0 & 64.8  & 13.3 &  64.8 & &468.0  & 48.0 & 48.0 &  20.0  \\ 
 6   & 77.4  & 80.8 & 73.1 & 64.8  & 15.0 &  64.8 & &526.5  & 48.0 & 48.0 &  20.0  \\ 
 7   & 86.0  & 80.8 & 81.3 & 64.8  & 16.7 &  64.8 & &585.0  & 48.0 & 48.0 &  20.0  \\ 
8    & 103.2  & 80.8 & 97.5 & 64.8  & 20.0 &  64.8 & &702.0  & 48.0 & 48.0 &  20.0 \\ 
 9   & 114.7  & 80.8 & 108.3 & 64.8  & 22.2 &  64.8 & &780.0  & 48.0 & 48.0 &  20.0\\ 
10   & 129.0  & 80.8 & 121.9 & 64.8  & N/A & N/A  &  &N/A   & N/A & N/A & N/A  \\ 
11   & 143.4  & 80.8 & 135.4 & 64.8  & N/A & N/A  &  &N/A   & N/A & N/A & N/A \\ \hline 
     & \multicolumn{6}{c|}{64 stations IEEE 802.11 ax} &  &\multicolumn{4}{c|}{ 4 stations IEEE 802.11 ac} \\ \hline
 0   & 3.8   & 136.8 & 3.5   & 64.8  & 0.8  &  64.8 & &58.5   & 48.0 & 48.0 & 20.0 \\ 
 1   & 7.5  & 136.8 & 7.1  & 64.8  & 1.7  &  64.8 & &117.0  & 48.0 & 48.0 & 20.0 \\ 
 2   & 11.3  & 100.8 & 10.6  & 64.8  & 2.5  &  64.8 & &175.5  & 48.0 & 48.0 & 20.0 \\ 
 3   & 15.0  & 100.8 & 14.2  & 64.8  & 3.3  &  64.8 & &234.0  & 48.0 & 48.0 & 20.0 \\ 
 4   & 22.5  & 88.8 & 21.3  & 64.8  & 5.0  &  64.8 & &351.0  & 48.0 & 48.0 & 20.0 \\ 
 5   & 30.0  & 88.8 & 28.3  & 64.8  & 6.7 &  64.8 & &468.0  & 48.0 & 48.0 & 20.0 \\ 
 6   & 33.8  & 88.8 & 31.9  & 64.8  & 7.5 &  64.8 & &526.5  & 48.0 & 48.0 & 20.0 \\ 
 7   & 37.5  & 88.8 & 35.4  & 64.8  & 8.3 &  64.8 & &585.0  & 48.0 & 48.0 & 20.0 \\ 
 8   & 45.0  & 88.8 & 42.5  & 64.8  & 10.9 &  64.8 & &702.0  & 48.0 & 48.0 & 20.0\\ 
 9   & 50.0  & 88.8 & 47.2  & 64.8  & 11.1 &  64.8 & &780.0  & 48.0 & 48.0 & 20.0\\ 
10   & N/A & N/A & N/A &  N/A & N/A &  N/A &  &N/A   & N/A & N/A & N/A \\ 
11   & N/A & N/A & N/A &  N/A & N/A &  N/A &  &N/A   & N/A & N/A & N/A \\ \hline 
\end{tabular}  
\end{table}

Concerning non-legacy transmissions, 
we assume a GI of $0.8 \mu s$ for transmissions over the DL.
For transmissions over the UL we assume a GI of $1.6 \mu s$. 
Therefore, the OFDM symbols are of $13.6 \mu s$ and $14.4 \mu s$ 
over the DL and the UL respectively.
Regarding legacy transmissions, the OFDM symbols are $4 \mu s$.

We assume the
Best Effort Access Category
in which $AIFS=43 \mu s$, $SIFS=16 \mu s$ and $CW_{min}=16$
for the transmissions of the AP.
The BackOff interval is a random number
chosen uniformly from
the range $[0,....,CW_{min}-1]$. 
Since we consider a very `large' number
of transmissions from the AP and we assume
that there are no collisions, we take the BackOff average
value of  
$\ceil{\frac{CW_{min}-1}{2}}$ and the average BackOff interval is
$\ceil{\frac{CW_{min}-1}{2}} \cdot SlotTime$
which equals $67.5 \mu s$ for a $SlotTime= 9 \mu s$. 
We also assume that the MAC Header is of 28 bytes and
the FCS is of 4 bytes.
We use the above values for the various parameters
since these are the default  ones suggested by the WiFi Alliance~\cite{WIFI}.

Finally, we consider several channel conditions which
are expressed by different values of the Bit Error Rate (BER)
which is the probability that a bit arrives corrupted
at the destination. We assume a model where these probabilities
are bitwise independent~\cite{L1}.

\section{Throughput analysis}

Let $X$ be the number
of MPDU frames in an A-MPDU frame, numbered $1,..,X$, 
and $Y_i$ is the number
of MSDUs in MPDU number $i$. 
Let $MacHeader$, $MpduDelimiter$ and $FCS$ be the length,
in bytes, of the MAC Header, MPDU Delimiter and FCS
fields respectively, and let $O_M = MacHeader+MpduDelimiter+FCS$.
Let $L_{DATA}$ be the length, in bytes,
of the MSDU frames.
Also, let 
$Len=4 \cdot \ceil{\frac{L_{DATA}+14}{4}}$ 
and $C_i = 8 \cdot 4 \cdot \ceil{\frac{O_M+Y_i \cdot Len}{4}}$.
$C_i$ is the length, in bits, of MPDU number $i$.

In the entire analysis ahead we assume that the Ack frames'
transmissions are all successful because Ack frames are short and
in most cases are transmitted in legacy mode.

\subsection{Single User mode}

The throughput in both 11ax and 11ac for
the traffic pattern in Figure~\ref{fig:traffic}(A)
is given by
Eq.~\ref{equ:thrtwole}~\cite{SA} where BER is the Bit Error Rate:

\begin{equation}
Thr=
\frac
{\sum_{i=1}^{X} 8 \cdot Y_i \cdot L_{DATA} \cdot (1-BER)^{C_i}}
{AIFS+BO(average)+P_{DL}+ T(DATA)+SIFS+P_{UL} +T(BAck)}
\label{equ:thrtwole}
\end{equation}

where:

\begin{eqnarray}
T(DATA)= TSym_{DL} \cdot \ceil{\frac{\sum_{i=1}^{X} C_i + 22}{TSym_{DL} \cdot R_{DL}}}
\\ \nonumber
T(BAck) = TSym_{UL} \cdot \ceil{\frac{(30 \cdot 8) +22}{TSym_{UL} \cdot R_{UL}}}
\\ \nonumber
\label{equ:timeac}
\end{eqnarray}

\normalsize

The term $BO(average)$ refers to the average value of the BackOff
interval, as given in Section 3.5. As was explained in 
Section 3.5 we use an average
value for this interval since 
there are no collisions.

$T(DATA)$ and $T(BAck)$ are the transmission times
of the data A-MPDU frames and BAck frames respectively.
$T(BAck)$
is based on the BAck frame's
lengths given in Figure~\ref{fig:frameformat}.
When assuming 30 bytes we consider
the acknowledgment of 64 MPDUs in the BAck.

$TSym_{DL}$ and $TSym_{UL}$ are
the lengths of the OFDM symbols on
the DL and the UL respectively, and every transmission
must be of an integral number of OFDM symbols.
The additional 22 bits in the numerators of $T(DATA)$ and $T(BAck)$
are due to the SERVICE and TAIL fields added to every
transmission by the PHY layer conv. protocol~\cite{IEEEBase1}.
$R_{DL}$ and $R_{UL}$ are the DL and UL PHY rates respectively
and $P_{DL}$ and $P_{UL}$ are the preambles used in
the DL and in the UL respectively (see Figure~\ref{fig:formatPPDU}).
 
The term in Eq.~\ref{equ:thrtwole} is not continuous, so
it is difficult to find the optimal $X$ and $Yi(s)$, i.e.
the values for $X$ and $Yi(s)$ that maximize 
the throughput. However, in~\cite{SA} it is
shown that if one neglects the rounding in the
denominator of Eq.~\ref{equ:thrtwole} then the optimal
solution has the property that all the MPDUs
contain almost the same number of MSDUs: the difference
between the largest and smallest number of MSDUs in MPDUs
is at most 1. The difference is indeed 1
if the limit on transmission time of the PPDU
does not enable transmission of
the same number of MSDUs in all MPDUs.

If neglecting the rounding of the denominator 
of Eq.~\ref{equ:thrtwole}, the received throughput
for every $X$ and $Y$ (Y is the equal number of MSDUs
in MPDUs) is as large as that received
in Eq.~\ref{equ:thrtwole}. The difference depends
on denominator size.

We therefore use the result in~\cite{SA} and look for the
maximum throughput as follows: We check for every
$X$, $1 \le X \le 64$ (also $1 \le X \le 256$ for 11ax)
and for every $Y$, $1 \le Y \le Y_{max}$, what is the
received throughput such that $Y_{max}$ is the
maximum possible number of MSDUs in an MPDU.
All is computed taking into account
the upper limit of $5.484 ms$ on the transmission time
of the PPDU (PSDU+preamble). If it is not possible
to transmit the same number of MSDUs in all
the MPDUs, part of the MPDUs have one more MSDU
than the others, up to the above upper limit
on the transmission time. We found that the smallest
denominator of any of the maximum throughputs
is around $1000 \mu s$. Neglecting the rounding
in the denominator
reduces its size by at most $2 \cdot 13.6 \mu s$ in 11ax
and $2 \cdot 4 \mu s$ in 11ac. Thus, the mistake in the
received maximum throughputs is
at most 2.8$\%$.

\subsection{Multi User mode}

The throughputs of 11ac and 11ax are 
given in Eq.~\ref{equ:thrac}-\ref{equ:axback} and 
their derivation can be found in~\cite{SA}.

\noindent
The throughput of 11ac for the traffic
pattern in Figure~\ref{fig:traffic}(B)
is given in Eq.~\ref{equ:thrac}:

\small

\begin{equation}
Thr_{AC} = 
\frac{4 \cdot \sum_{i=1}^{X} 8 \cdot Y_i \cdot L_{DATA} \cdot (1-BER)^{C_i}}
{AIFS+BO(average)+P_{DL}+T(DATA)+7 \cdot (SIFS + P_{UL}) + 4 \cdot 
T(BAck) + 3 \cdot T(BAR)}
\label{equ:thrac}
\end{equation}

\normalsize

\noindent
where:

\begin{eqnarray}
T(DATA)= TSym_{DL} \cdot \ceil{\frac{\sum_{i=1}^{X} C_i + 22}{TSym_{DL} \cdot R_{DL}}}
\\ \nonumber
T(BAck) = TSym_{UL} \cdot \ceil{\frac{(30 \cdot 8) +22}{TSym_{UL} \cdot R_{UL}}}
\\ \nonumber
T(BAR) = TSym_{UL} \cdot \ceil{\frac{(24 \cdot 8) +22}{TSym_{UL} \cdot R_{UL}}}
\label{equ:timeax}
\end{eqnarray}

\noindent
are the transmission times of the data A-MPDU frames, the BAck frames
and the BAR frames respectively. The transmission times
of the BAck and BAR frames are based on their
lengths given in Figure~\ref{fig:frameformat}.
$R_{DL}$ is the DL PHY rate and $R_{UL}$ is the UL PHY rate.
We have the multiplier of 4 in the numerator of Eq.~\ref{equ:thrac} since
the AP transmits simultaneously
to 4 stations. Also, $P_{DL}$ and $P_{UL}$ are
the lengths of the preambles in the DL and in the UL
respectively and $TSym_{DL}$ and $TSym_{UL}$ are the
lengths of the OFDM symbols used in the DL and UL respectively.

\noindent
The throughput of 11ax for the traffic pattern
in Figure~\ref{fig:traffic}(C)
is given in Eq.~\ref{equ:thrax}:

\small

\begin{equation}
Thr_{AX} = 
\frac{S \cdot \sum_{i=1}^{X} 8 \cdot Y_i \cdot L_{DATA} \cdot (1-BER)^{C_i}}
{AIFS+BO(average)+P_{DL}+T^{'}(DATA)+ PE + SIFS + P_{UL} + T^{'}(BAck) + PE}
\label{equ:thrax}
\end{equation}

\normalsize

\noindent
where:

\begin{eqnarray}
\label{equ:axback}
T^{'}(DATA)= TSym_{DL} \cdot \ceil{\frac{\sum_{i=1}^{X} C_i + ((O_M+72) \cdot 8) + 22}{TSym_{DL} \cdot R_{DL}}}
\\ \nonumber
T^{'}(BAck) = TSym_{UL} \cdot \ceil{\frac{(30 \cdot 8) +22}{TSym_{UL} \cdot R_{UL}}}
\end{eqnarray}

\noindent
$P_{DL}$ and $P_{UL}$ are again the preambles in
the DL and UL respectively.

In the term for $T^{'}(DATA)$ we assume the case of a Trigger Frame
which holds for $X$ data MPDUs in the A-MPDU frame such that
$19 \le X \le 64$. For $1 \le X \le 18$ it is more
efficient to use the HE Control Element of 4 bytes
added to every data MPDU, and the term $((O_M+72) \cdot 8)$
is therefore replaced by $(X \cdot 4 \cdot 8)$.
Notice that the 72 bytes come from 33 bytes of the TF frame, 
28 bytes of the MAC Header, 4 bytes of the FCS field,
4 bytes of the MPDU Delimiter and rounding to an integral
number of 4 bytes.
For the BAck frame, $T^{'}(BAck)$
is based on a BAck frame acknowledging 64 MPDUs.
In 11ax it is also possible to acknowledge
256 MPDUs and in this case the 30 bytes in
$T^{'}(BAck)$ are replaced by 54 bytes.
See Figure~\ref{fig:frameformat}(B).
Notice the multiplier $S$ in the
numerator of Eq.~\ref{equ:thrax}. $S$ is either
4, 8, 16, 32 or 64, the number
of stations to which the AP transmits simultaneously.

Again, the terms in Eqs.~\ref{equ:thrac} and~\ref{equ:thrax}
are not continuous and 
therefore we again use
the result in~\cite{SA},
as in the SU mode,
and look for the
maximum throughput as specified in Section 4.1 .

The analytical results of 11ax have been verified by 
an 11ax simulation model running on the $ns3$
simulator~\cite{NS3} and the simulation and analytical
results are the same. This outcome is not surprising however, because
there is not any stochastic process involved
in the scheduled transmissions in 11ax
assumed in this paper. 
Therefore, we do not mention the simulation results any further
in this paper.

\section{An approximation of the optimal A-MPDU structure}

In this section we show an approximation to the value
of $X_{OPT}$, the number of optimal MPDUs in an A-MPDU, i.e.
the number of MPDUs that maximizes the throughput, as a function
of the BER. We concentrate on 11ax although the computation is
valid for 11ac as well.

\subsection{The case BER$>$0}

We re-write Eq.~\ref{equ:thrax} by ignoring the rounding
of $T^{'}(DATA)$ and $T^{'}(BAck)$, ignoring the 22 bits in
the numerators of $T^{'}(DATA)$ and $T^{'}(BAck)$, settings
$O_p = AIFS+BO+SIFS+P_{UL}+T^{'}(BAck)+PE$, assuming that every
MPDU has the same number $Y$ of MPDUs,
$O_M = MacHeader+MpduDelimiter+FCS$ and ignoring the overhead
due to the TF frame:

\begin{equation}
Thr = \frac
{S \cdot X \cdot Y \cdot 8 \cdot L_{DATA} \cdot (1-BER)^{8 \cdot (Y \cdot Len +O_M)}}
{O_p +P_{DL} + 
\frac{X \cdot 8 \cdot (Y \cdot Len +O_M)}{R_{DL}}
\label{equ:appen1}
}
\end{equation}

Notice that given a number $Y$ of MPDUs in an A-MPDU,
the throughput increases as $X$ increases. Therefore,
it is worthwhile
to transmit as large A-MPDUs as possible, up to the limit
on the transmission time of the A-MPDU frame. Let $T$ be this
limit, $5484 \mu s$ in our case. Then, the following 
approximation on the relation between $X$ and $Y$ can be written:

\begin{equation}
T=
\frac
{X \cdot 8 \cdot (Y \cdot len + O_M)}
{R_{DL}}
+ P_{DL}
\label{equ:appen2}
\end{equation}

\noindent
or: 

\begin{equation}
X= \frac
{R_{DL} \cdot (T-P_{DL})}
{8 \cdot (Y \cdot Len + O_M)}
\label{equ:appen3}
\end{equation}

In Eqs.~\ref{equ:appen2} and~\ref{equ:appen3} we approximate
that the sum of the A-MPDU transmission time plus the DL preamble
is $T$.

We now substitute the term for X in Eq.~\ref{equ:appen1} by the term in
Eq.~\ref{equ:appen3} and receive:

\begin{equation}
Thr = \frac
{S \cdot 
\frac{R_{DL} \cdot (T-P_{DL})}{8 \cdot (Y \cdot Len + O_M)}
\cdot Y \cdot 8 \cdot L_{DATA} \cdot (1-BER)^{8 \cdot (Y \cdot Len + O_M)}}
{T+O_p-P_{DL}}
\label{equ:appen4}
\end{equation}

Notice that the denominator of Eq.~\ref{equ:appen4} is a constant
and so to find the maximum throughput as a function of Y one needs
to find the maximum of the following function:

\begin{equation}
\frac{Y}{8 \cdot (Y \cdot Len + O_M)} \cdot
(1-BER)^{8 \cdot (Y \cdot Len + O_M)}
\label{equ:appen5}
\end{equation}

\noindent
The optimal $Y$, $Y_{OPT}$, is given in Eq.~\ref{equ:appen6}:

\begin{equation}
Y_{OPT}=
\frac
{O_M \cdot (\sqrt{1-\frac{4}{8 \cdot O_M \cdot ln(1-BER)}}-1)}
{2 \cdot Len}
\label{equ:appen6}
\end{equation}

\noindent
Notice that by Eq.~\ref{equ:appen3} we can now write the optimal X,
$X_{OPT}$, as:

\begin{equation}
X_{OPT}=
\frac
{R_{DL} \cdot (T-P_{DL})}
{8 \cdot O_M 
(\frac
{(\sqrt{1-\frac{4}{8 \cdot O_M \cdot ln(1-BER)}}-1)}
{2 \cdot Len}
+1)
}
\label{equ:appen7}
\end{equation}

\noindent
Notice that we look for an integer $Y_{OPT}$ and that $Y_{OPT}$ must
be at least 1. Therefore, Eq.~\ref{equ:appen7} is only an approximation
for $X_{OPT}$.

Consider now Figure~\ref{fig:res3}(F) as an example (we
refer to Figure~\ref{fig:res3} more deeply later). We have for this case
$P_{DL}=88.8 \mu s$, $R_{DL}=50 Mbps$ and $O_M = 36$ bytes. We also
have three cases of $Len$, $Len=1516, 528$ and $80$ bytes for
MSDUs of lengths $1500, 512$ and $64$ bytes respectively.
For all three cases we receive that $Y_{OPT}= \frac{653}{Len}$.
For $Len=1516, 528$ and $80$ bytes we receive
$Y_{OPT}= 0.43, 1.23$ and $8.16$ respectively.
For $Y_{OPT}=0.43$ we need to round up to 1 and receive
$X_{OPT}=21.72$. It turns out that
$X_{OPT}=21$ yields a larger throughput than 22 MPDUs.
For $Y_{OPT}=1.23$ we can take either $\floor{Y_{OPT}}=1$
or $\ceil{Y_{OPT}}=2$.
For the two cases we receive $\floor{X_{OPT}}=59$ and $30$
respectively where the first case yields a larger throughput.
We handle the case for $Len=80$ similarly, where the $X_{OPT}$
is now 50. All these values for $X_{OPT}$ appear in Figure~\ref{fig:res3}(F).

\begin{figure}
\vskip 9cm
\includegraphics{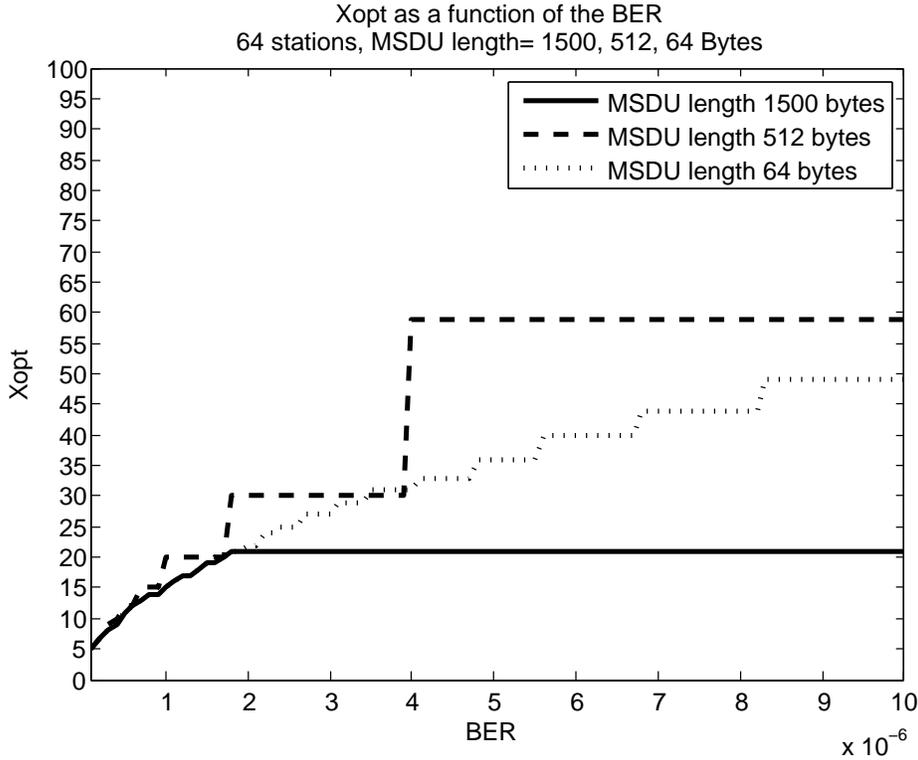}
\caption{$X_{OPT}$ as a function of the BER and MSDU length, 64 stations, in IEEE 802.11ax .}
\label{fig:appendix1}
\end{figure}

In Figure~\ref{fig:appendix1} we plot three curves
for the values of $X_{OPT}$ as a function of the BER
for MSDUs of 1500, 512 and 64 bytes respectively.
Notice that for an MSDU of 1500 bytes 21 MPDUs of 1 MSDU
is the optimal number of MPDUs over a wide range of BER
values. This is because as the BER increases it is worthwhile
transmitting short MPDUs, but one MSDU must
be included in an MPDU. For MSDUs of 512 bytes
there is more flexibility in the number of MSDUs per MPDU and so the
optimal number of MPDUs is more flexible. For MSDUs of 60 bytes the number
of MSDUs per MPDU varies according to the BER in the most
flexible way and so does the number of MPDUs.
The number of optimal MPDUs is smaller than
in MSDUs of 512 bytes because the smaller size of the MSDUs
enables using the MPDUs more efficiently, the MPDUs are
little longer than in the case of 512 bytes MSDUs and due to the
limit on the A-MPDU transmission time, a smaller number of MPDUs 
is needed.

\subsection{The case BER$=$0}

For BER$=$0 Eq.~\ref{equ:appen1} becomes:

\begin{equation}
Thr = \frac
{S \cdot X \cdot Y \cdot 8 \cdot L_{DATA} }
{O_p +P_{DL} + 
\frac{X \cdot 8 \cdot (Y \cdot Len +O_M)}{R_{DL}}
\label{equ:appen8}
}
\end{equation}

\noindent
and one needs to optimize the function:

\begin{equation}
\frac
{Y}
{8 \cdot (Y \cdot Len +O_M)}
\label{equ:appen9}
\end{equation}

\noindent
which reveals that in every MPDU it is worthwhile to contain
the maximum number of MPDUs, $Y_{MAX}$, 
which is $\floor{\frac{11454-O_M}{Len}}$.

\noindent
Therefore:

\begin{equation}
X_{OPT}=
\frac
{R_{DL} \cdot (T-P_{DL})}
{8 \cdot ( \floor{\frac{11454-O_M}{Len}} \cdot Len + O_M)}
\label{equ:appen10}
\end{equation}

\noindent
For example, for Figure~\ref{fig:res3}(D) we have
$R_{DL}=50 Mbps$, $P_{DL}=88.8 \mu s$, $O_M=36$ bytes and
$X_{OPT}= \frac{33720}{\floor{\frac{11418}{Len}} \cdot Len + 36}$

For MSDUs of 1550, 512 and 64 bytes one receives
$Len=1516, 528$ and $80$ bytes respectively, which
gives $X_{OPT}=3.166, 3.031, 2.958$ respectively. 
Since we look for an integer $X_{OPT}$ one needs to
choose between 3 or 4 MPDUs for
the first two cases and between 2 or 3 MPDUs for the
third case. It turns out that 3,3,3 are the optimal
number of MPDUs respectively, as appears
in Figure~\ref{fig:res3}(D).

\section{Throughput's models and results}

\subsection{Transmissions' models and scenarios}

We compare between all applicable configurations
and DL service scheduling
flavors of the AP transmissions 
to up to 64 stations. The service scheduling flavors are
as follows:

\noindent
Concerning 11ac :

\begin{itemize}

\item
DL SU, UL SU Back transmission in legacy mode, up to 64 MPDUs 
in an A-MPDU frame, denoted previously as $SU_{AC}(1)$.

\item
DL 4 users MU-MIMO, UL 4 times SU BAck transmission in legacy mode, 
up to 64 MPDUs in an A-MPDU frame, denoted previously as $MU_{AC}(4)$.

\end{itemize}

\noindent
Concerning 11ax :

\begin{itemize}

\item
DL SU, UL SU BAck transmission in legacy mode, up to 64 or 256 MPDUs 
in an A-MPDU frame, 
denoted previously as 11ax/64 and 11ax/256 $SU_{AX}(1)$ respectively.

\item
DL 4 users MU-MIMO, UL MU-MIMO or OFDMA BAck transmission, 
up to 64 or 256 MPDUs in an A-MPDU frame, denoted previously as 11ax/64
and 11ax/256 $MU_{AX}(4)$ respectively.

\item
DL S=8, 16, 32, 64 users DL MU-MIMO + OFDMA,
UL MU-MIMO+OFDMA or OFDMA BAck transmission, 
up to 64 or 256 MPDUs in an A-MPDU frame, 
denoted previously as 11ax/64 and 11ax/256 $MU_{AX}(S)$ respectively.

\end{itemize}

For every number $S$ of stations we analyze the optimal
DL service scheduling
working point, i.e. the one that optimizes throughput,
as a function of the transmission flavor,
MCS in use and the A-MPDU frame structure.

First, we checked for every
number of stations all possible transmission 
DL service scheduling flavors that are applicable
for this number of stations. For example, for 64 stations one can
use 64 cycles of Figure~\ref{fig:traffic}(A) sequentially both
in 11ac and 11ax, i.e. 
$64 \cdot SU_{AC}(1)$ or $64 \cdot SU_{AX}(1)$. 
One can also use 16 cycles of Figures~\ref{fig:traffic}(B)
and~\ref{fig:traffic}(C) in 11ac and 
11ax respectively, namely $16 \cdot MU_{AC}(4)$
and $16 \cdot MU_{AX}(4)$ respectively.
Finally, one can also use 8, 4, 2 and 1 cycles of Figure~\ref{fig:traffic}(C)
in 11ax, denoted previously
$MU_{AX}(8)$, $MU_{AX}(16)$, $MU_{AX}(32)$ 
and $MU_{AX}(64)$
respectively. 

Every transmission flavor is checked over all
applicable MCSs. For 11ac these are MCS0-MCS9. For 11ax
these are MCS0-MCS11 except in the
case of 64 stations, where only MCS0-MCS9 are
applicable. We also check for every transmission flavor
and MCS the optimal working point by optimizing
the number of MPDUs and number of MSDUs
in every MPDU that yields the maximum throughput, i.e. we look
for the optimal A-MPDU frame structure.
We checked all the above for MSDUs of 64, 512 and 1500 bytes
and BER$=$$0, 10^{-6}, 10^{-5}$.

In the next section we show three sets of results.
In Figure~\ref{fig:res1} we show the maximum throughputs
received for every number of stations in every transmission
flavor for MSDUs of 1500 bytes. The results for
MSDUs of 64 and 512 bytes are similar. In Figure~\ref{fig:res2}
we demonstrate for $MU_{AX}(4)$ and 
$MU_{AX}(64)$ the maximum
throughputs received in the various DL
service scheduling flavors of 11ax, as a function
of the MCSs. The maximum
among them is shown in Figure~\ref{fig:res1}. Figure~\ref{fig:res2}
shows the influence of the maximum number of MPDUs in an
A-MPDU frame, 64 or 256 on the received throughput, as well
as the influence of using UL MU-MIMO or UL OFDMA on the
received throughput. Finally, in Figure~\ref{fig:res3} we show
the influence of the number of MPDUs in an A-MPDU frame on the
received throughput in cases of 4 and 64 stations,
for BER$=$0, $10^{-6}$ and $10^{-5}$.

\subsection{Throughput results}

\begin{figure}
\vskip 16cm
\includegraphics{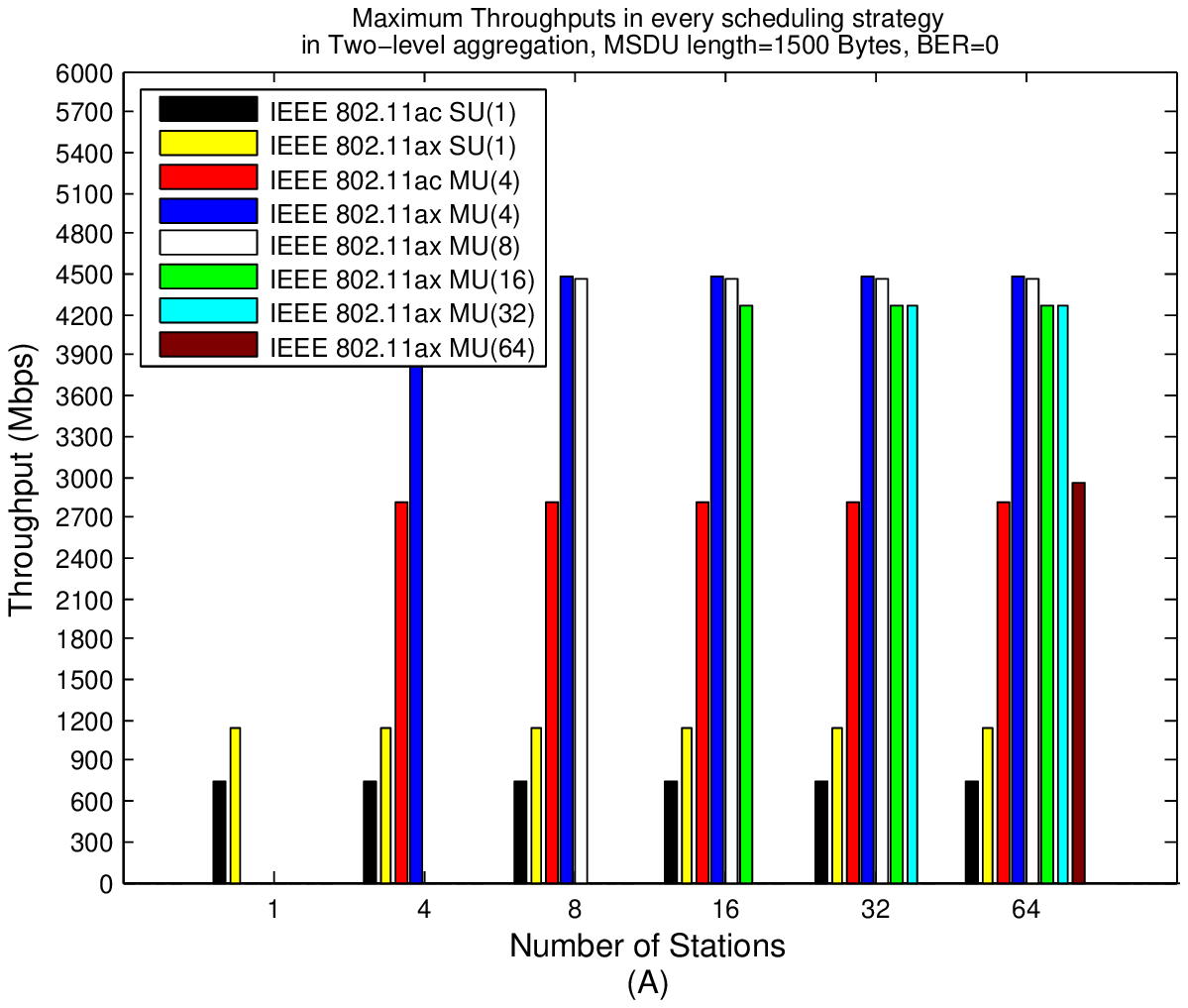}
\includegraphics{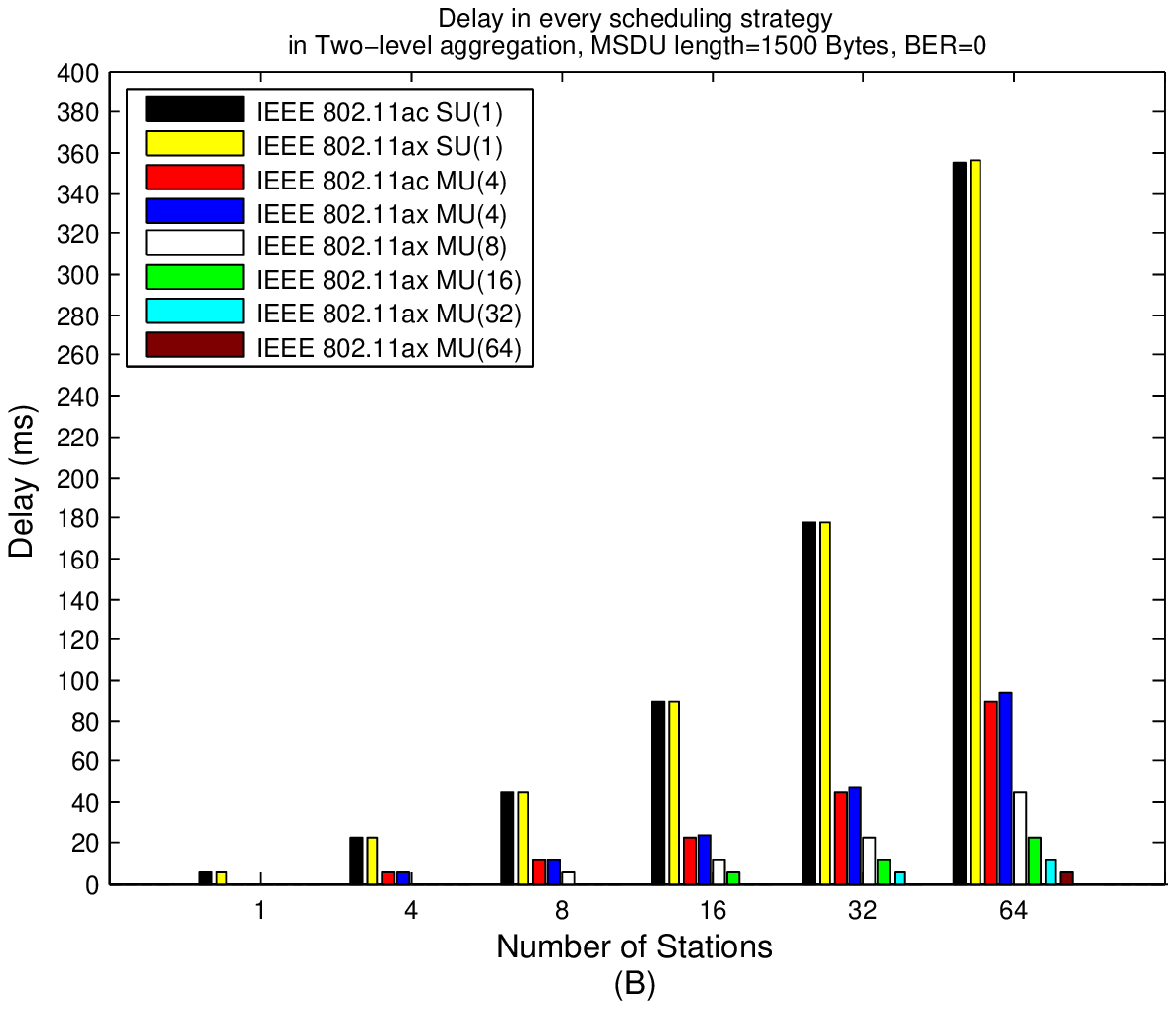}
\includegraphics{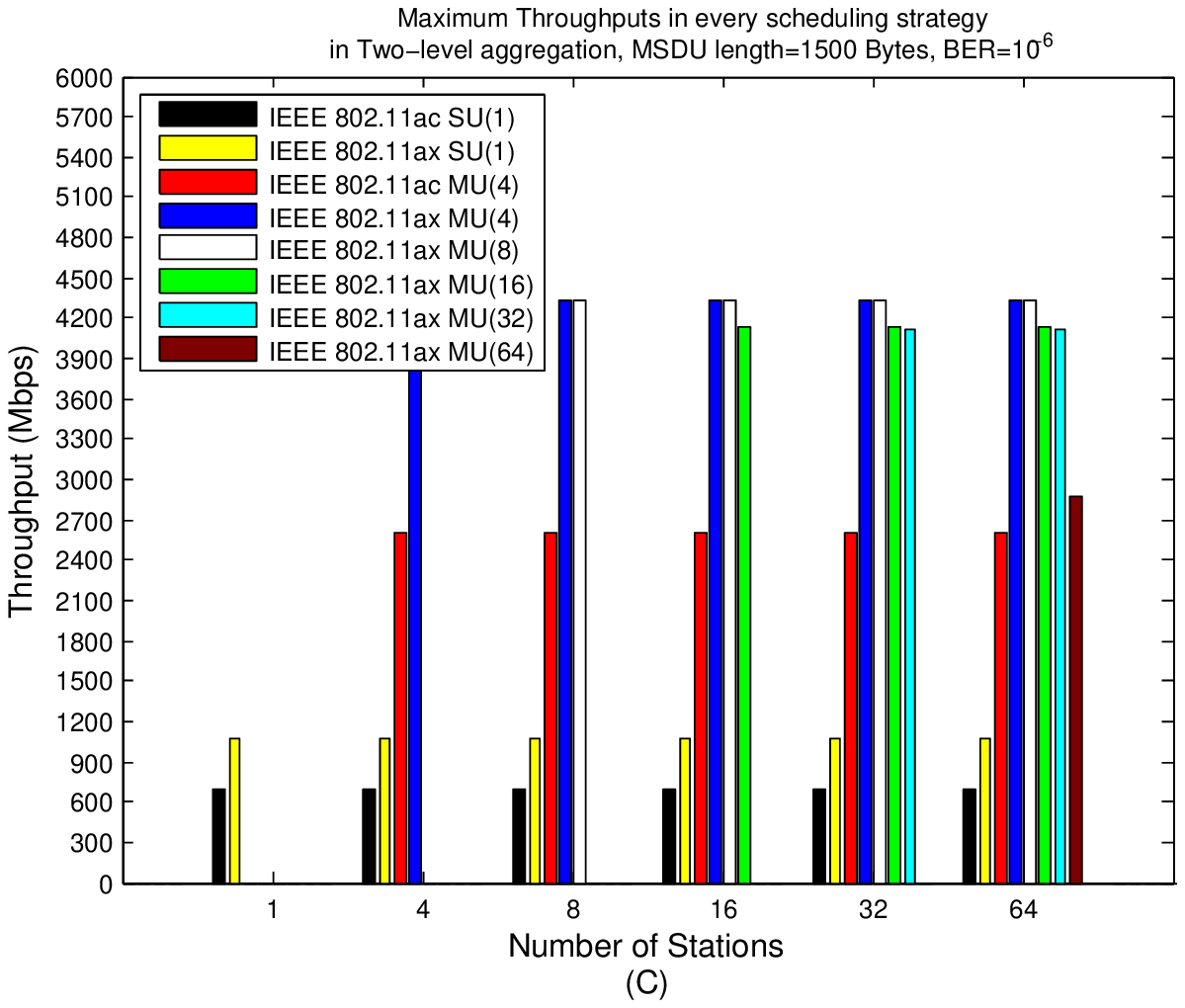}
\includegraphics{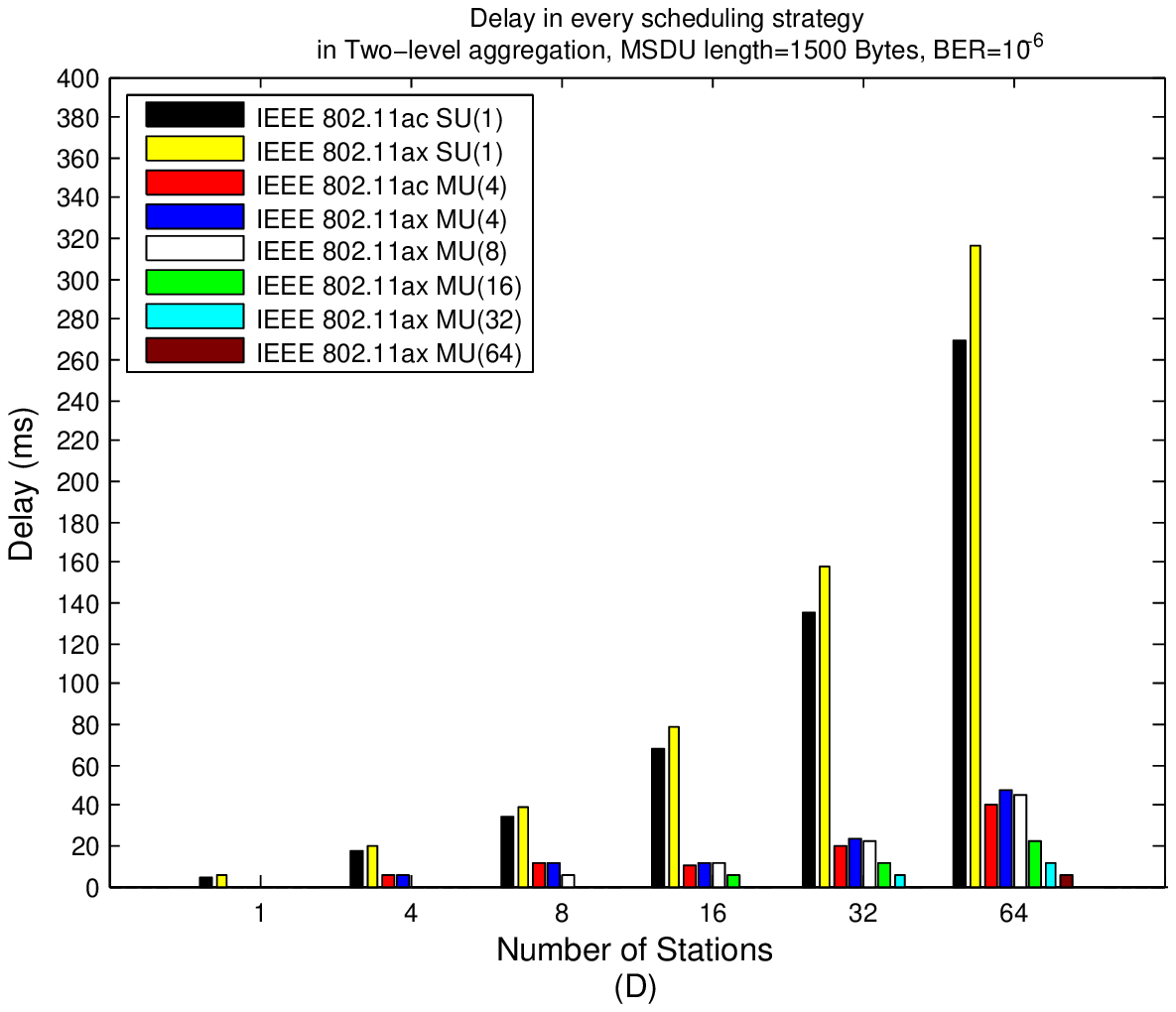}
\includegraphics{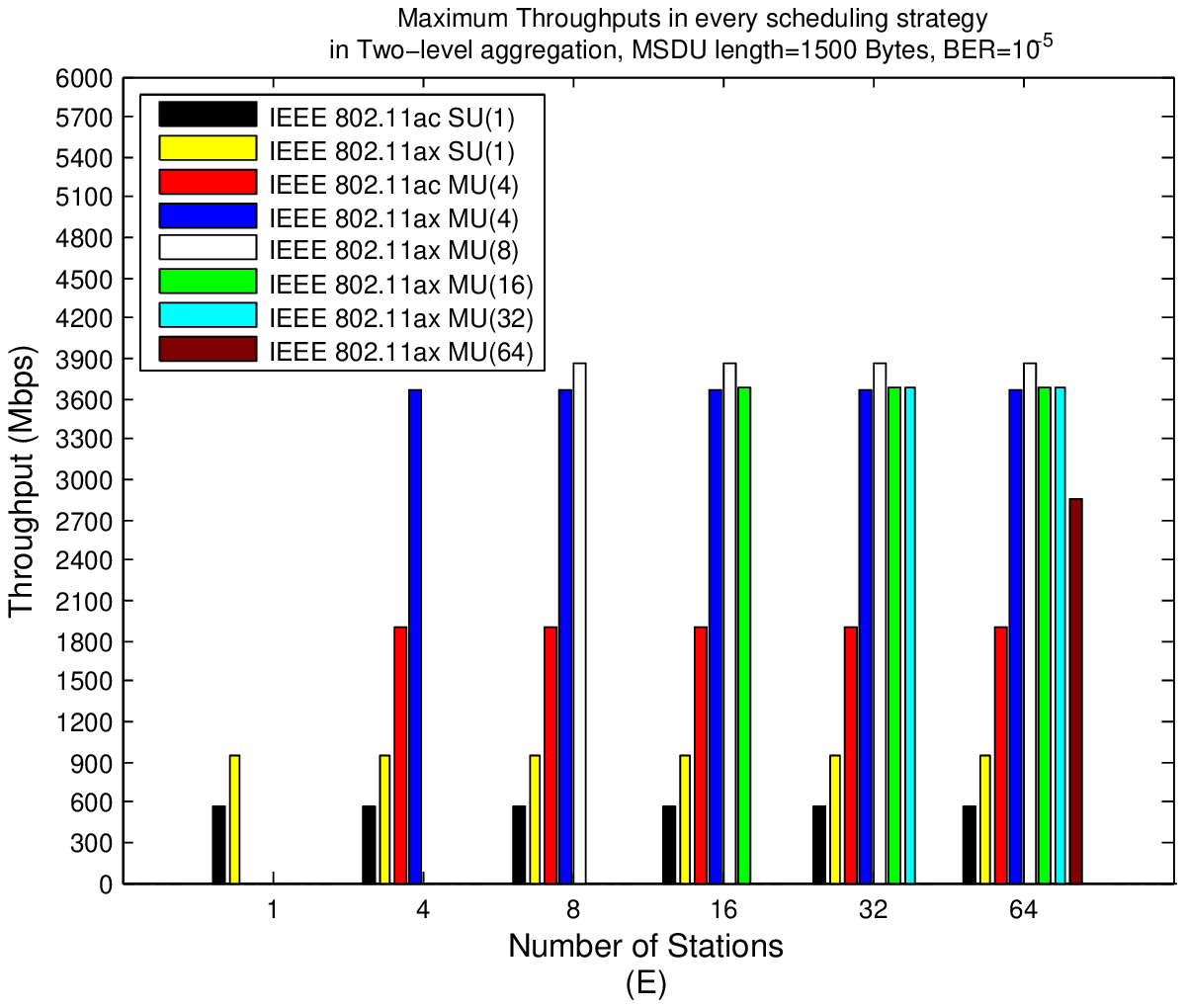}
\includegraphics{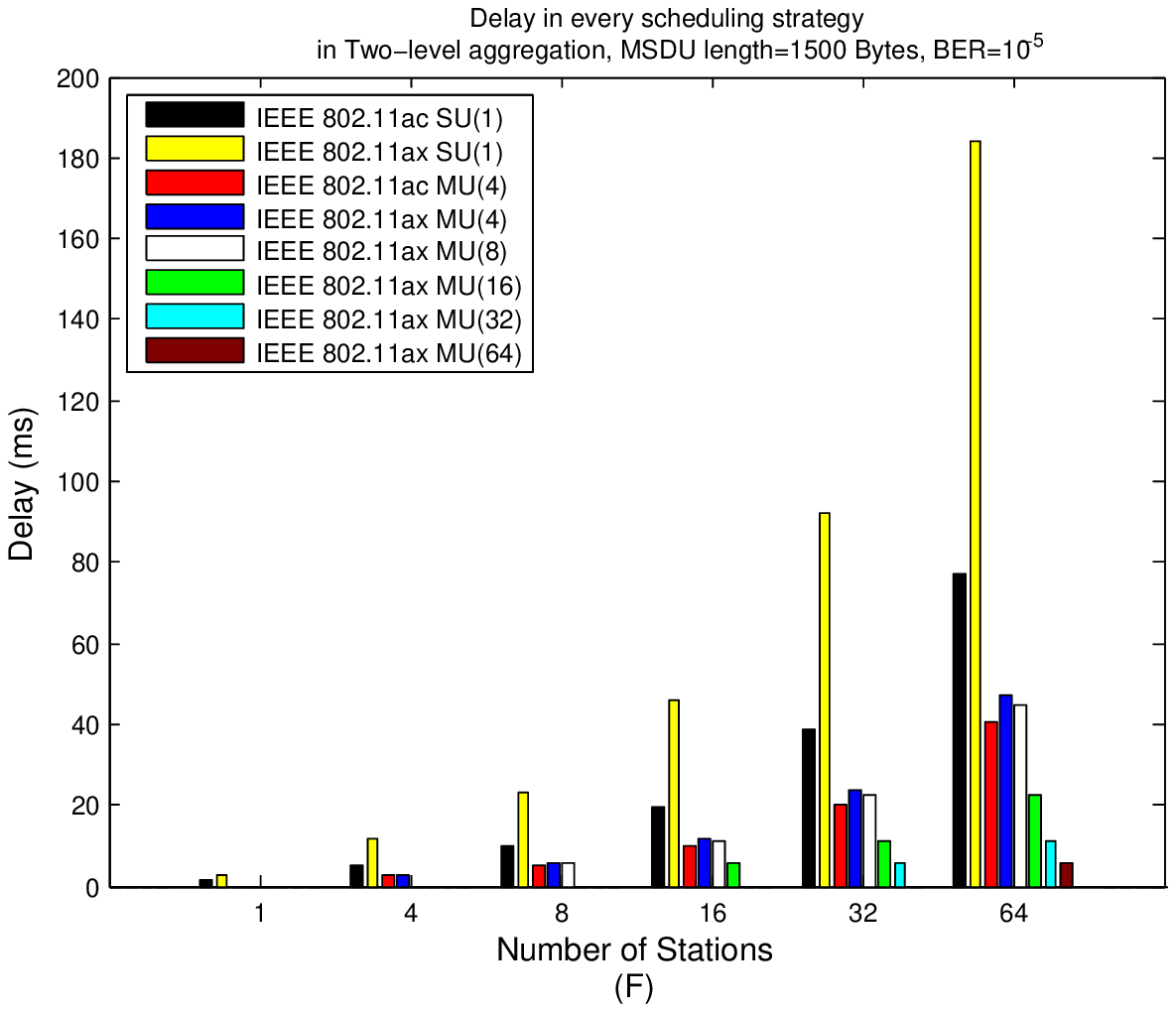}
\caption{Maximum throughputs and corresponding access delays in Single User and Multi User in IEEE 802.11ac and IEEE 802.11ax .}
\label{fig:res1}
\end{figure}

Recall that in Figure~\ref{fig:res1} we show the maximum throughputs
received as a function of the number
of stations to which the AP transmits. 
We show results for MSDUs of 1500 bytes only;
similar results are received for MSDUs
of 64 and 512 bytes. 

In Figure~\ref{fig:res1}(A) we show the results for
BER$=$0. When referring to e.g. 11ax MU(4) in the legend
we refer to $MU_{AX}(4)$, i.e. the 
case in which the AP transmits to 4 stations in 11ax
simultaneously using DL MU-MIMO, Figure~\ref{fig:traffic}(C).
When showing the results for $MU_{AX}(4)$ for
the case of e.g. 64 stations, the traffic
cycle in Figure~\ref{fig:traffic}(C) repeats
itself 16 times; every transmission is to a
different group of 4 stations, i.e. $16 \cdot MU_{AX}(4)$.

We see from Figure~\ref{fig:res1}(A) that the largest
throughput is received in $MU_{AX}(4)$. Notice that the
throughout of $MU_{AX}(8)$ is only slightly smaller
than that of $MU_{AX}(4)$. From Table 2 one can see that
the PHY rates in $MU_{AX}(8)$ are half of those of $MU_{AX}(4)$. This 
is balanced by twice the number of stations to which
the AP transmits. However, 
in $MU_{AX}(4)$ 522 MSDUs are transmitted in an A-MPDU
frame compared to 520 MSDUs in $MU_{AX}(8)$. Also,
the DL preamble in
$MU_{AX}(8)$ is slightly larger than in $MU_{AX}(4)$ 
due to the HE-SIG-B field.
These two factors reduce the throughput of
$MU_{AX}(8)$ compared to $MU_{AX}(4)$. 

In $MU_{AX}(16)$ the PHY rates are less than half
of those in $MU_{AX}(8)$ and together with the larger 
preamble this explains
why $MU_{AX}(16)$ has a smaller throughput 
than $MU_{AX}(8)$ and $MU_{AX}(4)$.
The explanation for the throughputs of $MU_{AX}(32)$ and $MU_{AX}(64)$
is similar to those given above for $MU_{AX}(8)$ and $MU_{AX}(16)$.
Notice that the PHY rates in $MU_{AX}(64)$ are less
than half of those of $MU_{AX}(32)$ and also that
MCS10 and MCS11 are not applicable for $MU_{AX}(64)$, which
is a main factor in the sharp decrease in the throughput of
$MU_{AX}(64)$ compared to $MU_{AX}(32)$.

Notice also that for all stations 11ax outperforms
11ac due to larger PHY rates and
simultaneous transmissions of BAck frames in the UL
compared to sequential transmissions in legacy mode in 11ac .
For 4, 8, 16, 32 and 64 stations and using MU-MIMO, 11ax
outperforms 11ac by $59\%$, 4470 vs. 2808 Mbps, the
throughputs in $MU_{AX}(4)$ and $MU_{AC}(4)$ respectively. In SU when
transmitting to 1 station only, 11ax outperforms 11ac
by $52\%$, 1133 vs. 742 Mbps.

Although the throughput metric is important, so is the access
delay metric, defined in this paper as the
time elapsed between two consecutive transmissions from
the AP to the same station.
Notice for example that in the
case of $MU_{AX}(4)$ that achieves the largest throughput,
the access delay in the case of 64 stations is 16 times
the cycle of Figure~\ref{fig:traffic}(C) while in $MU_{AX}(64)$
the access delay is only one such cycle.
Notice also that we refer here to the {\it access delay}
and not to the {\it packet delay}. 
Since there are retransmissions
in the IEEE 802.11 MAC, 
the packet delay
is defined as the delay since a packet is first transmitted
and until it is successfully received. 

In Figure~\ref{fig:res1}(B) we show the access delays
for the various DL service scheduling transmissions' flavors. 
Some applications benefit primarily from lower
latency, especially real-time streaming applications
such as voice, video conferencing or even video chat.
The trade-off between latency and throughput becomes
more complex as applications are scaled out to run 
in a distributed fashion. 
The access delay results are
as expected; the access delay is lower when the AP transmits
simultaneously to additional stations . It seems that the
cycles are about the same in length in all
DL service scheduling transmissions' flavors 
and the relation between
access delays is about the same between
the number of stations to which the AP transmits simultaneously.

In Figures~\ref{fig:res1}(C) and~\ref{fig:res1}(D) we show
the results for BER$=$$10^{-6}$. There are some trends
in this BER that become more prominent in 
BER$=$$10^{-5}$ so we concentrate now only on
BER$=$$10^{-5}$.

In Figure~\ref{fig:res1}(E) we show the maximum throughput
as a function of the number of stations for the case
BER$=$$10^{-5}$. An interesting difference compared to
BER$=$0 is that the best transmission flavor is $MU_{AX}(8)$
compared to $MU_{AX}(4)$ in BER$=$0. $MU_{AX}(8)$ outperforms
$MU_{AX}(4)$ due to the short MPDUs and its smaller
PHY rates. The optimal A-MPDU frame structure in both
DL service scheduling flavors 
is 255 MPDUs of one MSDU each. In $MU_{AX}(4)$ a cycle
lasts $2.944 ms$ and in $MU_{AX}(8)$ it is $5.583 ms$. 
In $MU_{AX}(8)$ twice the number of MSDUs are transmitted
than in $MU_{AX}(4)$, but this is done in less than twice
the cycle length of $MU_{AX}(4)$ due to equal overhead
in both DL service scheduling flavors.
This leads to a larger throughput in $MU_{AX}(8)$.
In BER$=$0 the cycle length of $MU_{AX}(4)$ is
$5.596 ms$ compared to $5.583 ms$ in $MU_{AX}(8)$, i.e.
about the same.
However, the number of MSDUs in $MU_{AX}(4)$ is
slightly larger than twice the number of MSDUs in $MU_{AX}(8)$
(522 vs. 520) and the
preamble is slightly shorter.
Therefore in BER$=$0 $MU_{AX}(4)$ has a slightly larger
throughput.

When comparing between the throughputs of $MU_{AX}(8)$ and
$MU_{AC}(4)$, 11ax outperforms 11ac by $103\%$ , 3872 vs 1902 Mbps
respectively. For SU(1) 11ax outperforms 11ac by $74\%$, 940
vs. 540 Mbps respectively.

In Figure~\ref{fig:res1}(F) we show the corresponding access
delays of the DL service scheduling
transmissions' flavors for BER$=$$10^{-5}$.
Notice that the access delay of $SU_{AX}(1)$ is much larger
than that of $SU_{AC}(1)$, in contrast to
BER$=$0 where they are about the same.
The difference is because the maximum throughput
of $SU_{AC}(1)$ is received when transmitting 64 MPDUs of 1 MSDU
each while in $SU_{AX}(1)$ the A-MPDU contains 256 MPDUs of
1 MSDU each. In BER$=$0 the MPDUs contain 7 MSDUs each, and in
both 11ac and 11ax the cycles are around $5.5 ms$. Therefore,
access delays are similar.

Also worth mentioning is the relation between the access delays
of $MU_{AX}(4)$ and $MU_{AX}(8)$. For BER$=$$10^{-5}$ they are about the same
because the maximum throughput in both
DL service scheduling flavors is 
received when an A-MPDU frame contains 255 MPDUs
of 1 MSDU each. Since the PHY rates in $MU_{AX}(8)$ are about
half of those in $MU_{AX}(4)$, the cycle length in $MU_{AX}(8)$ is
about double in length than in $MU_{AX}(4)$. However, this
is compensated by double the number of stations to which
the AP transmits in $MU_{AX}(8)$ compared to $MU_{AX}(4)$; 
overall the access delays are similar in both DL service
scheduling flavors.

In BER$=$0 the cycle length in both $MU_{AX}(4)$ and $MU_{AX}(8)$ are
about the same, around $5.5 ms$, transmitting as many MSDUs
as possible. The access delay in $MU_{AX}(4)$ is now twice than that
of $MU_{AX}(8)$ because of the 4 vs. 8 stations to which
the AP transmits in  $MU_{AX}(4)$ and $MU_{AX}(8)$ respectively.

Overall it can be concluded from Figure~\ref{fig:res1} that
there is not any one best flavor. For example,
$MU_{AX}(8)$ achieves
the maximum throughput 
but $MU_{AX}(16)$ and $MU_{AX}(32)$ also achieve high
throughput but with smaller access delays compared to $MU_{AX}(8)$.

\begin{figure}
\vskip 9cm
\includegraphics{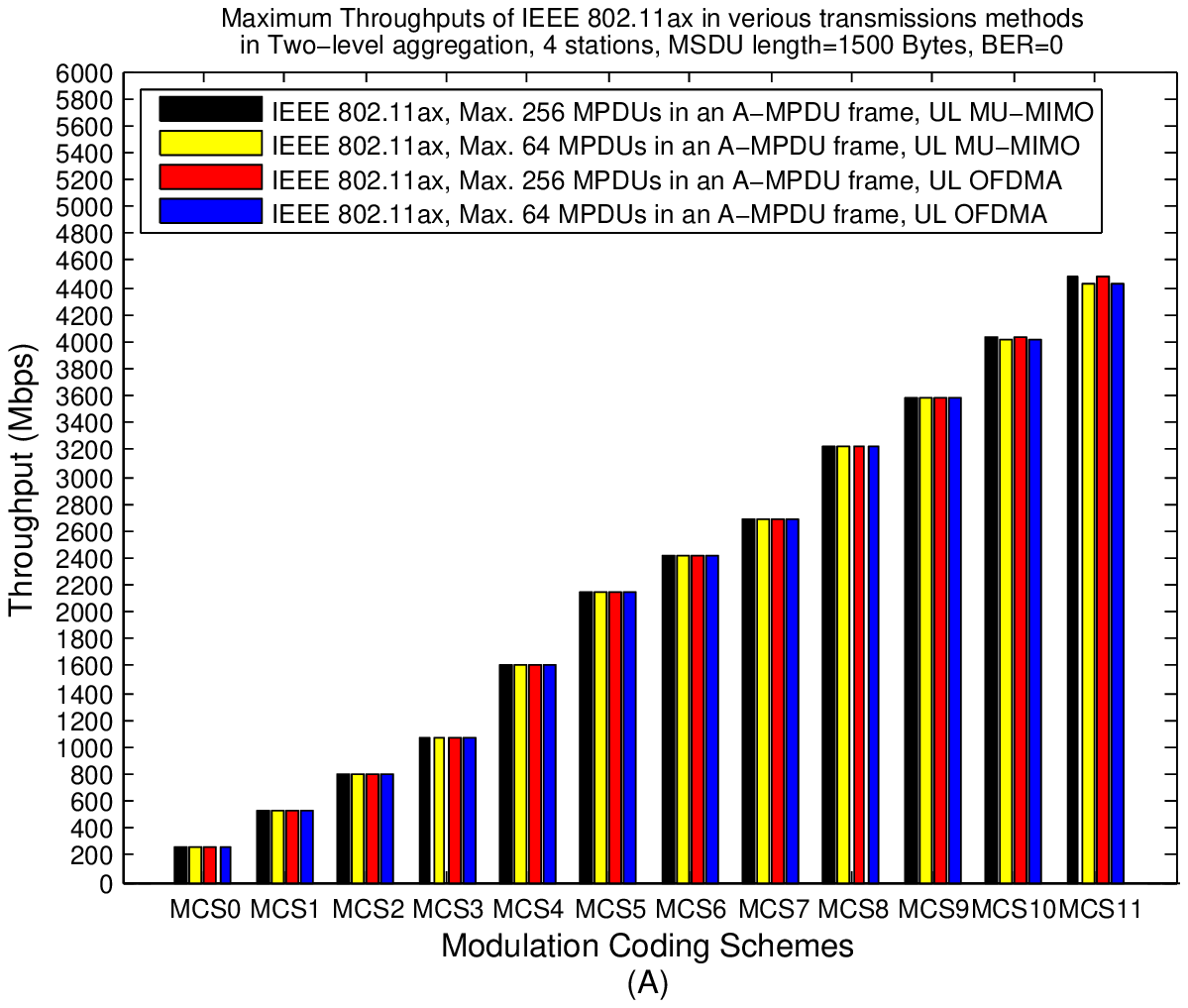}
\includegraphics{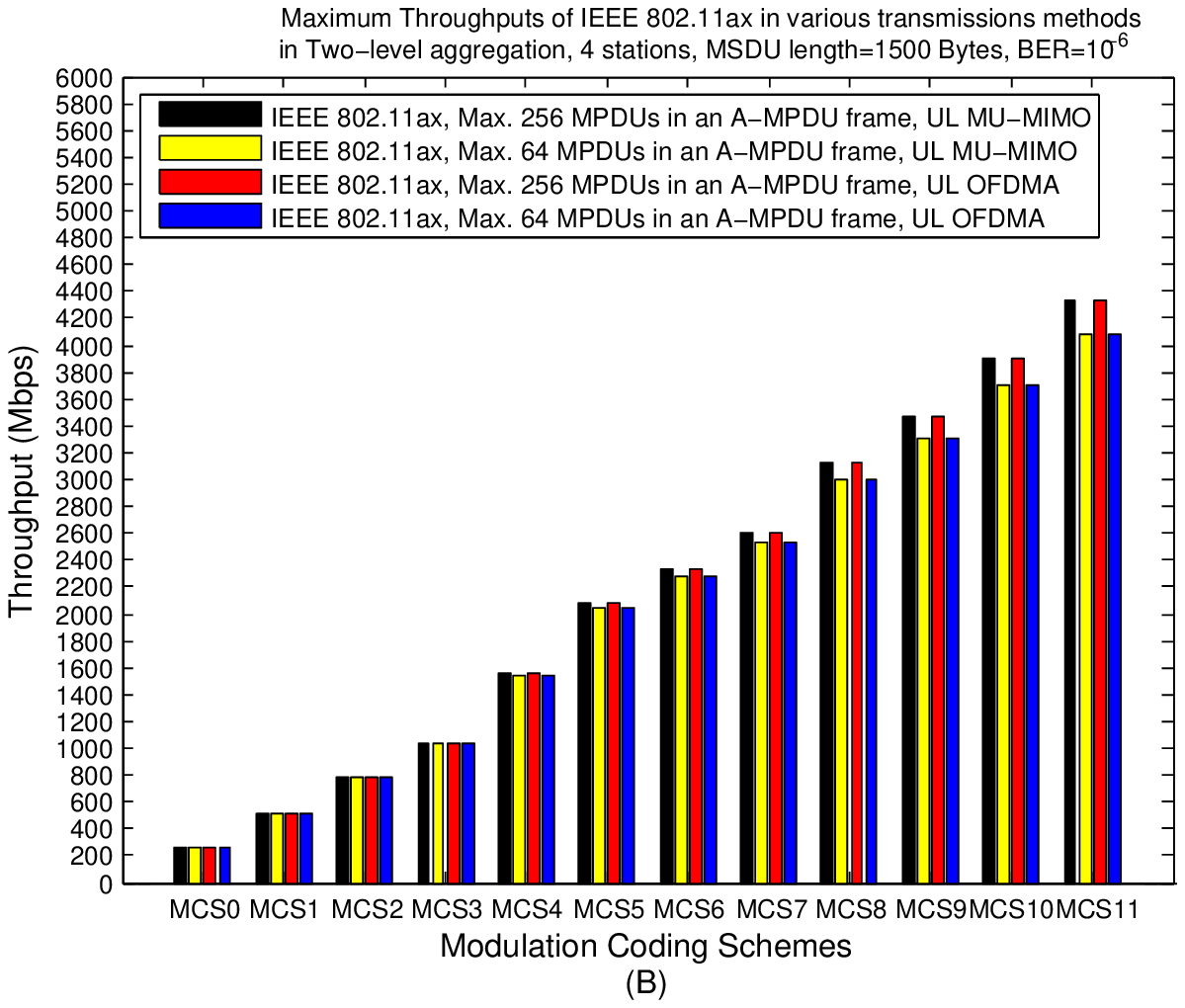}
\includegraphics{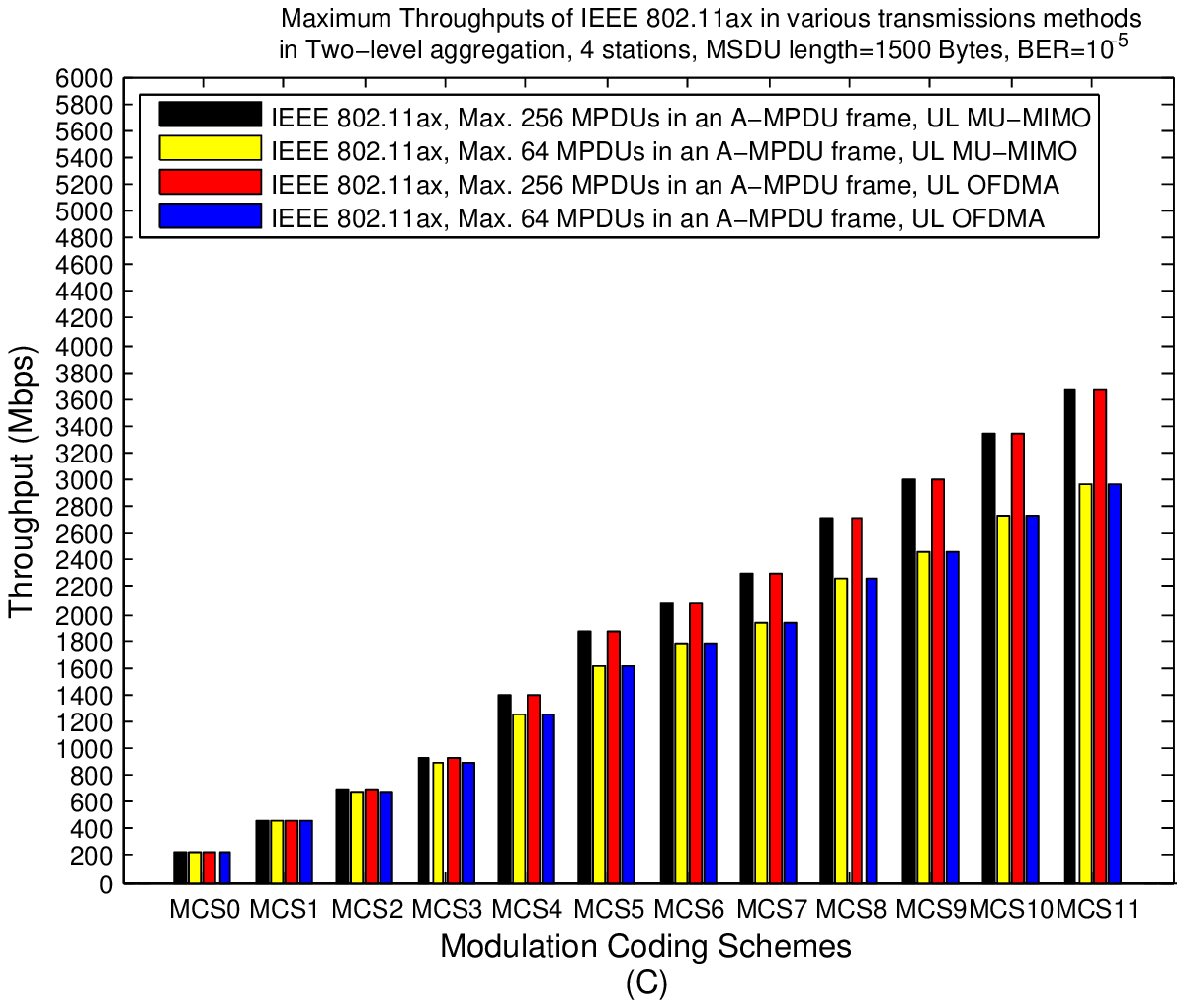}
\includegraphics{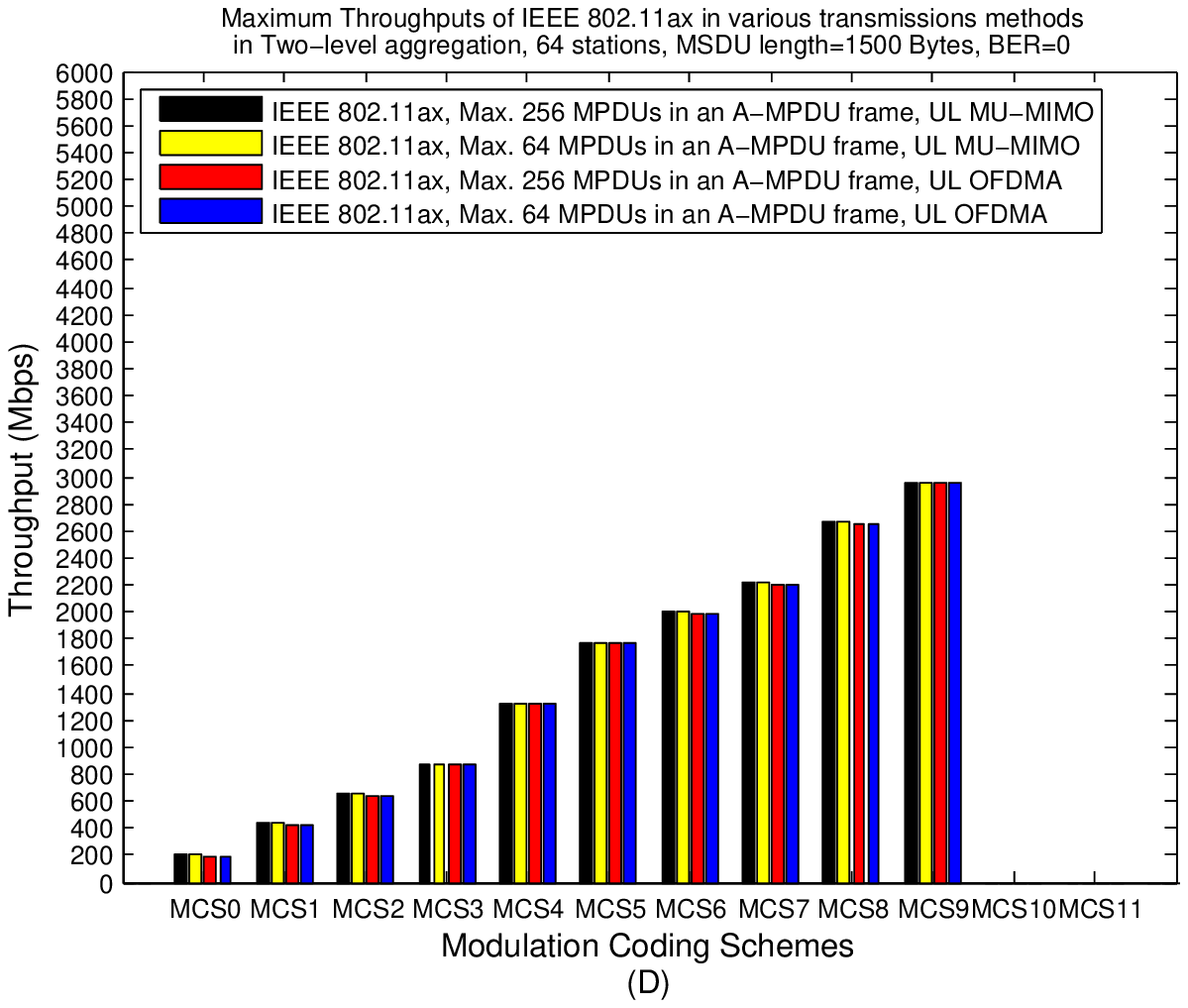}
\includegraphics{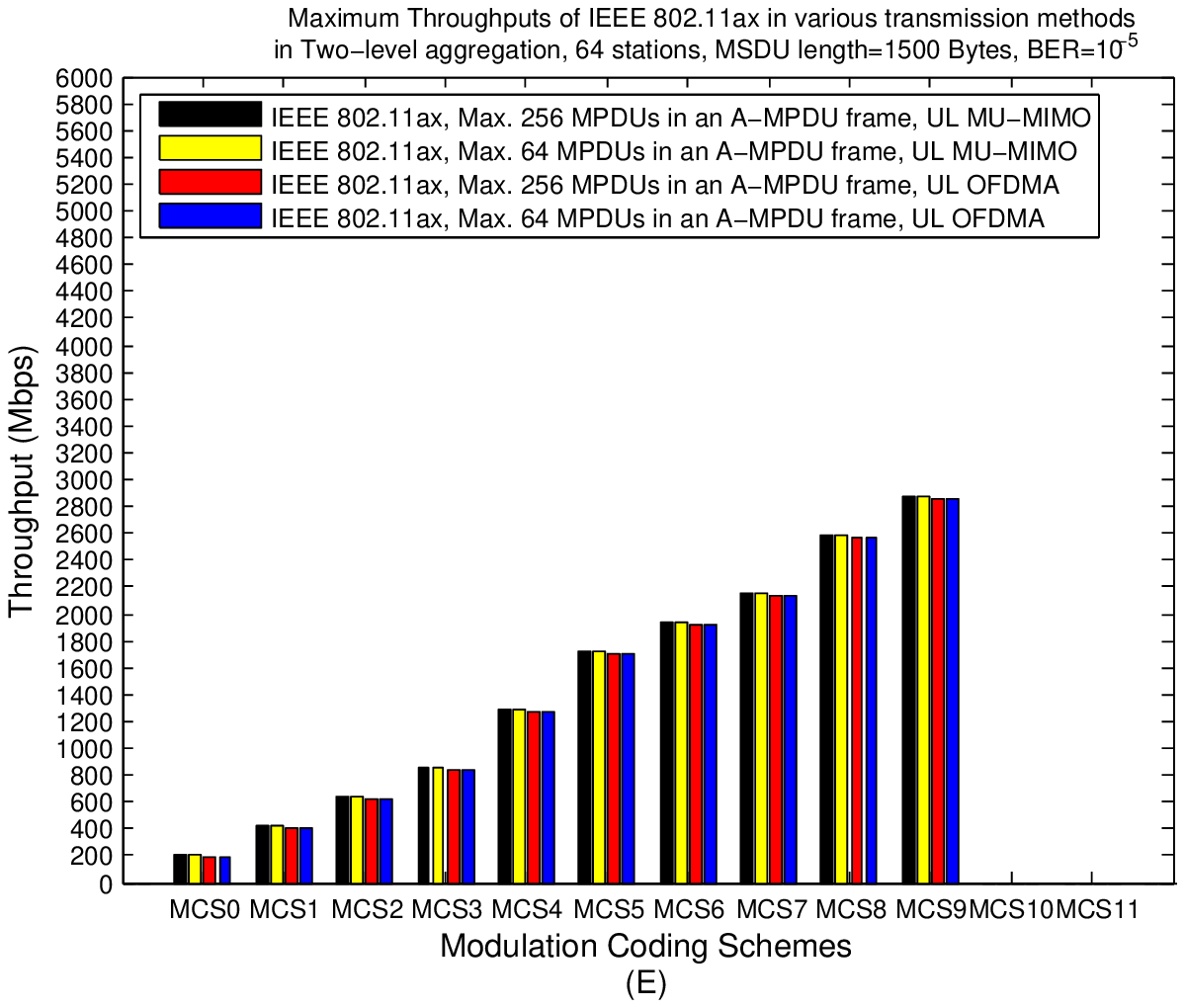}
\includegraphics{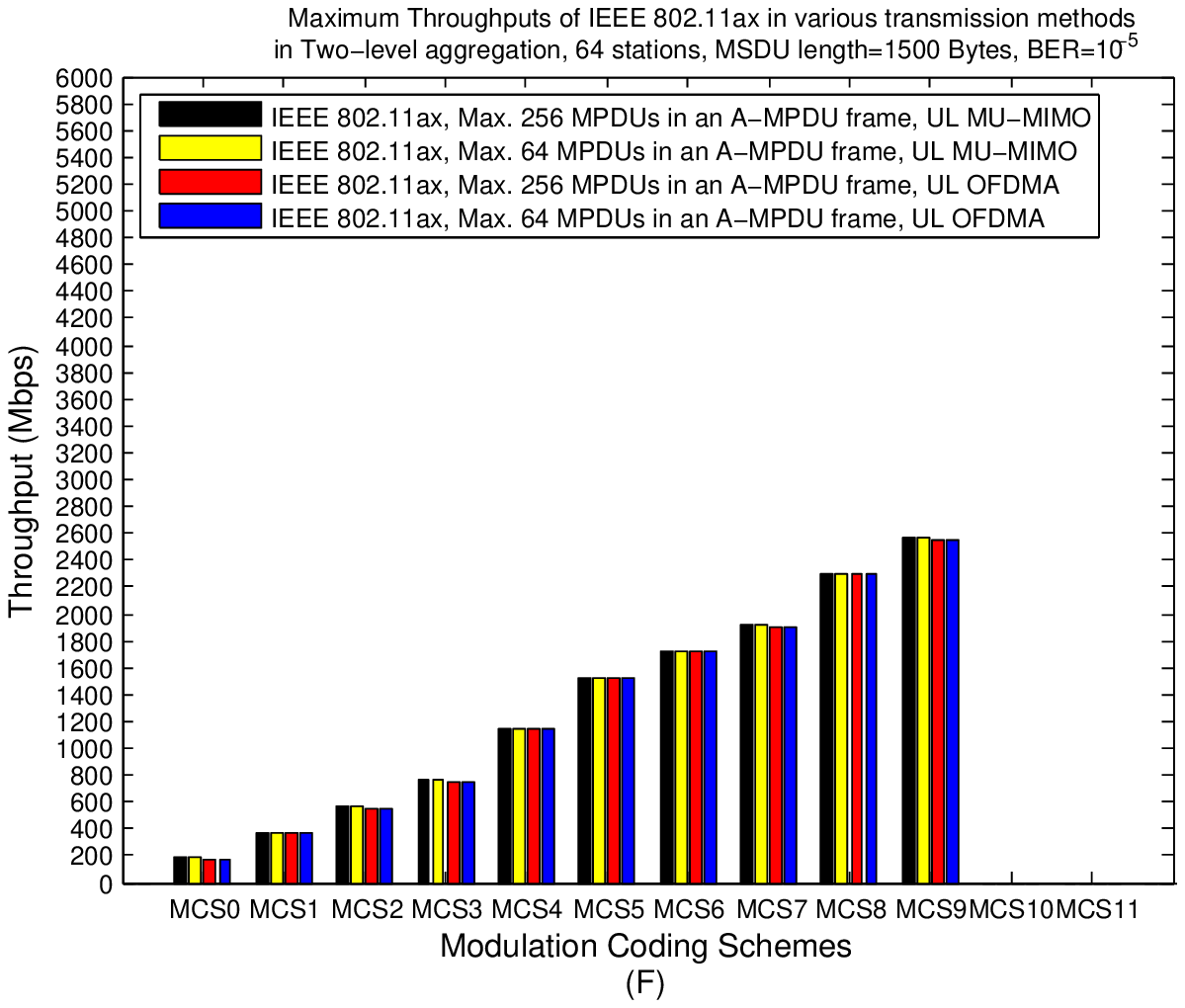}
\caption{The throughputs of the various transmissions methods in IEEE 802.11ax for Multi User when transmitting simultaneously to 4 and 64 stations.}
\label{fig:res2}
\end{figure}

In Figure~\ref{fig:res2} we show the throughput optimization performance
of $MU_{AX}(4)$ and $MU_{AX}(64)$ for every MCS, for the case
of UL MU-MIMO and UL OFDMA, 
for the cases using 64 and 256 MPDUs
in an A-MPDU frame 
and for
BER$=$0, $10^{-6}$ and $10^{-5}$. 
We again concentrate only on BER$=$$0, 10^{-5}$ because
the results for BER$=$$10^{-6}$ are similar in trend.
In Figures~\ref{fig:res2}(A) and~\ref{fig:res2}(C) we show
the results for $MU_{AX}(4)$ for BER$=$0 and BER$=$$10^{-5}$ respectively.
In Figures~\ref{fig:res2}(D) and~\ref{fig:res2}(F) the same results
are shown for $MU_{AX}(64)$.
Notice that for $MU_{AX}(64)$ there are no results for MCS10 and MCS11
which are not applicable in this case due to low
PHY rates.

The maximum throughput is always received in $MU_{AX}(4)$
in MCS11 ( MCS9 in $MU_{AX}(64)$ ) due to the
highest PHY rates in this MCS. Considering $MU_{AX}(4)$ notice
that for BER$=$0 11ax/256 outperforms 11ax/64 only in MCS10 and MCS11
while in BER$=$$10^{-5}$ 11ax/256 outperforms 11ax/64 starting
from MCS2 (starting from MCS5 in BER$=$$10^{-6}$).
In BER$=$0 it is efficient to transmit large MPDUs. Therefore,
the limit on the A-MPDU frame size is imposed by the limit
of $5.484 ms$ on the transmission time of the PPDU. Only
in larger PHY rates there is room for more than  64 MPDUs and
in these cases 11ax/256 has an advantage over 11ax/64 . In BER$=$$10^{-5}$
it is efficient to transmit short MPDUs. In this case the significant
limit is the number of MPDUs. 11ax/256 outperforms 11ax/64
from MCS2 because it enables transmitting more short MPDUs than
11ax/64 . A detailed analysis of this phenomenon
can be found in~\cite{SA1}.

Another interesting phenomenon is the relation between
UL MU-MIMO and UL OFDMA. When using UL OFDMA the UL PHY rates
are much smaller than those in UL MU-MIMO (see Table 2). However,
rounding $T^{'}(BAck)$ to an integral number
of OFDM symbols of $14.4 \mu$ ($12.8 \mu s + 1.6 \mu s$
Guard Interval) and the small size of
the BAck frames results in similar $T^{'}(BAck)$ times
in $MU_{AX}(4)$. In $MU_{AX}(64)$ the UL PHY rates
in UL OFDMA are even smaller and an additional
OFDM symbol is needed. Therefore, there is a slight 
advantage to UL MU-MIMO . This phenomenon 
is seen in Figure~\ref{fig:res2}(F)
where transmission to 64 stations is assumed.
Using DL MU-MIMO with up to 64 or 256 MPDUs
in the A-MPDU frame outperforms the same DL service scheduling
transmission
flavors respectively when using UL OFDMA. On the 
other hand this phenomenon is not seen in Figure~\ref{fig:res2}(C)
when transmitting to 4 stations.

\begin{figure}
\vskip 10cm
\includegraphics{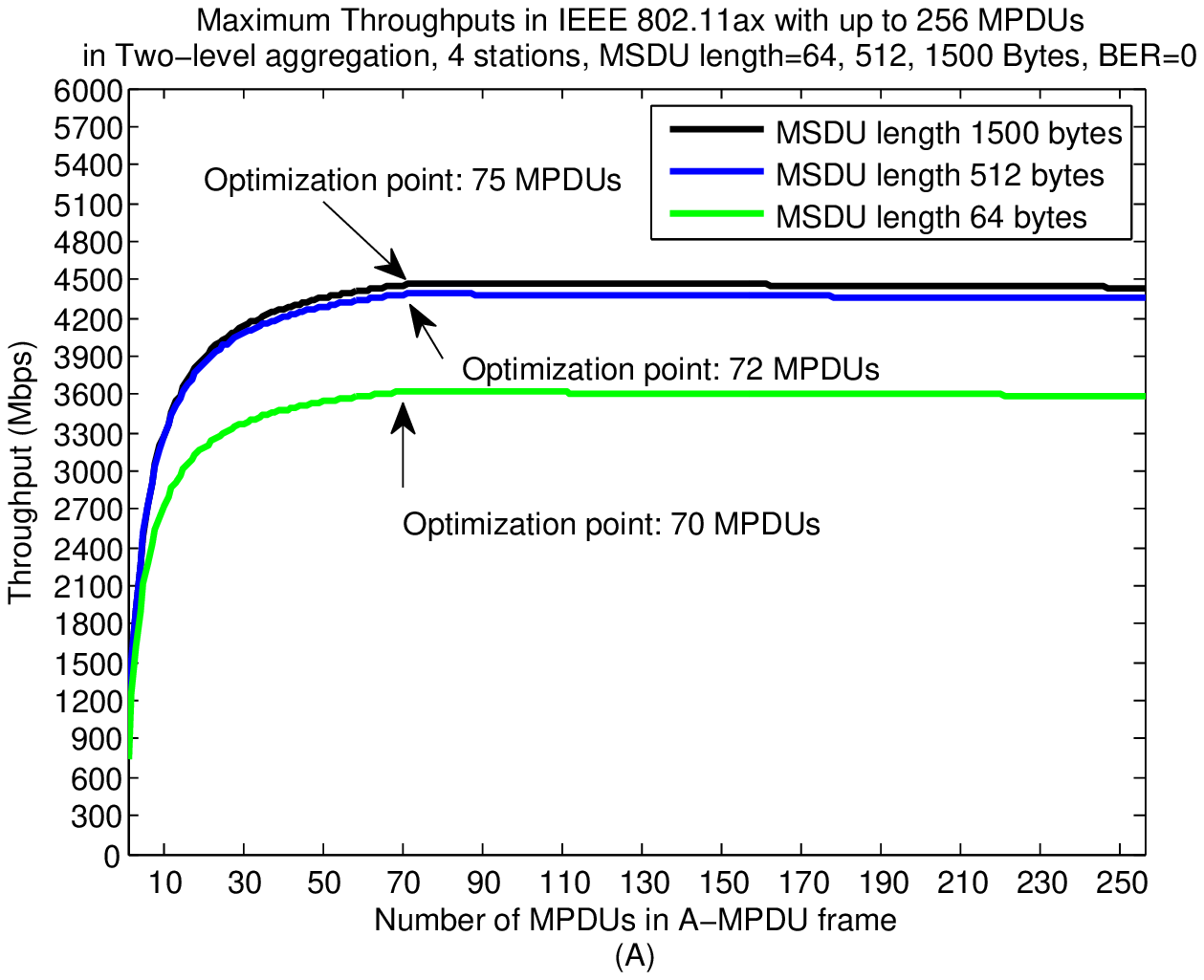}
\includegraphics{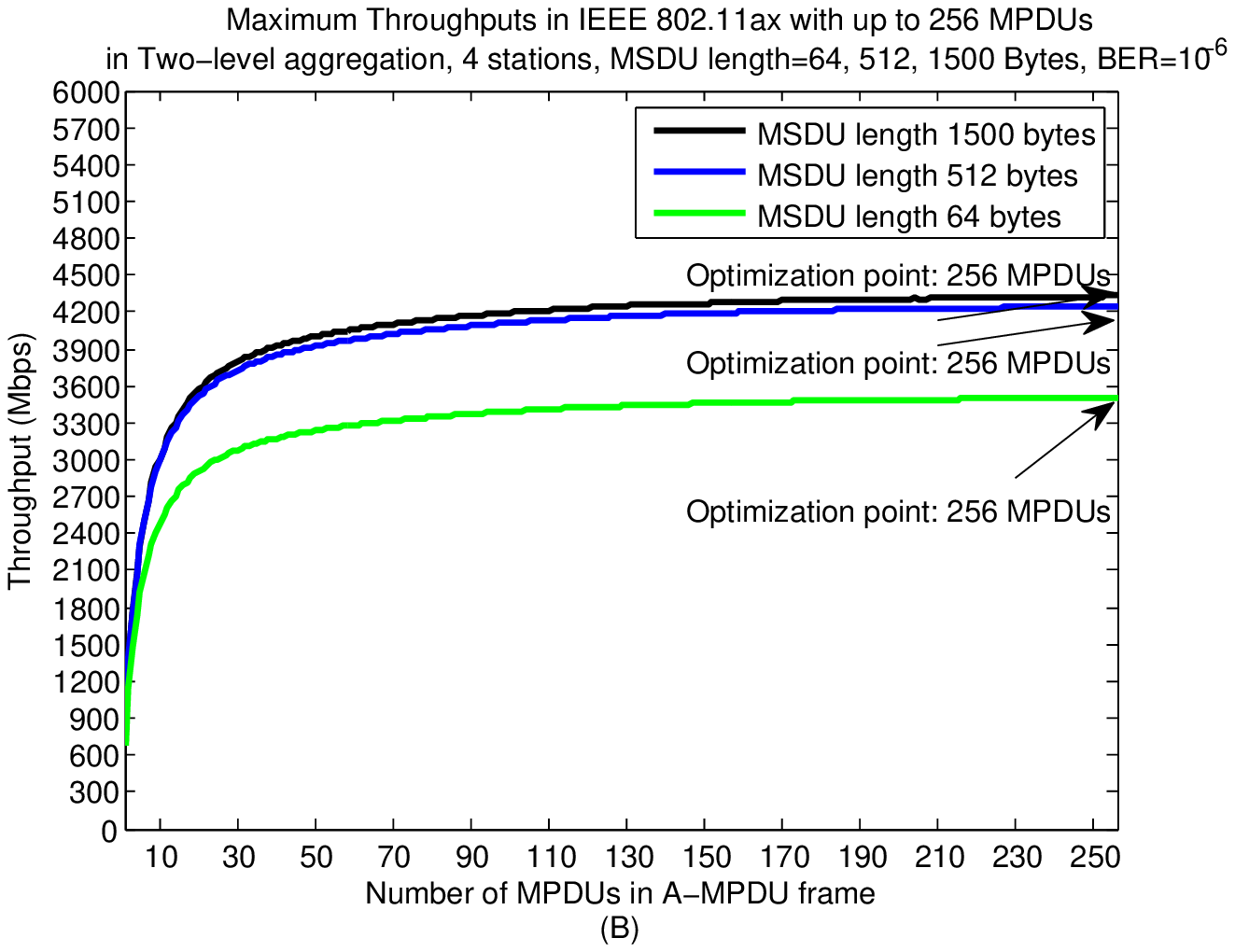}
\includegraphics{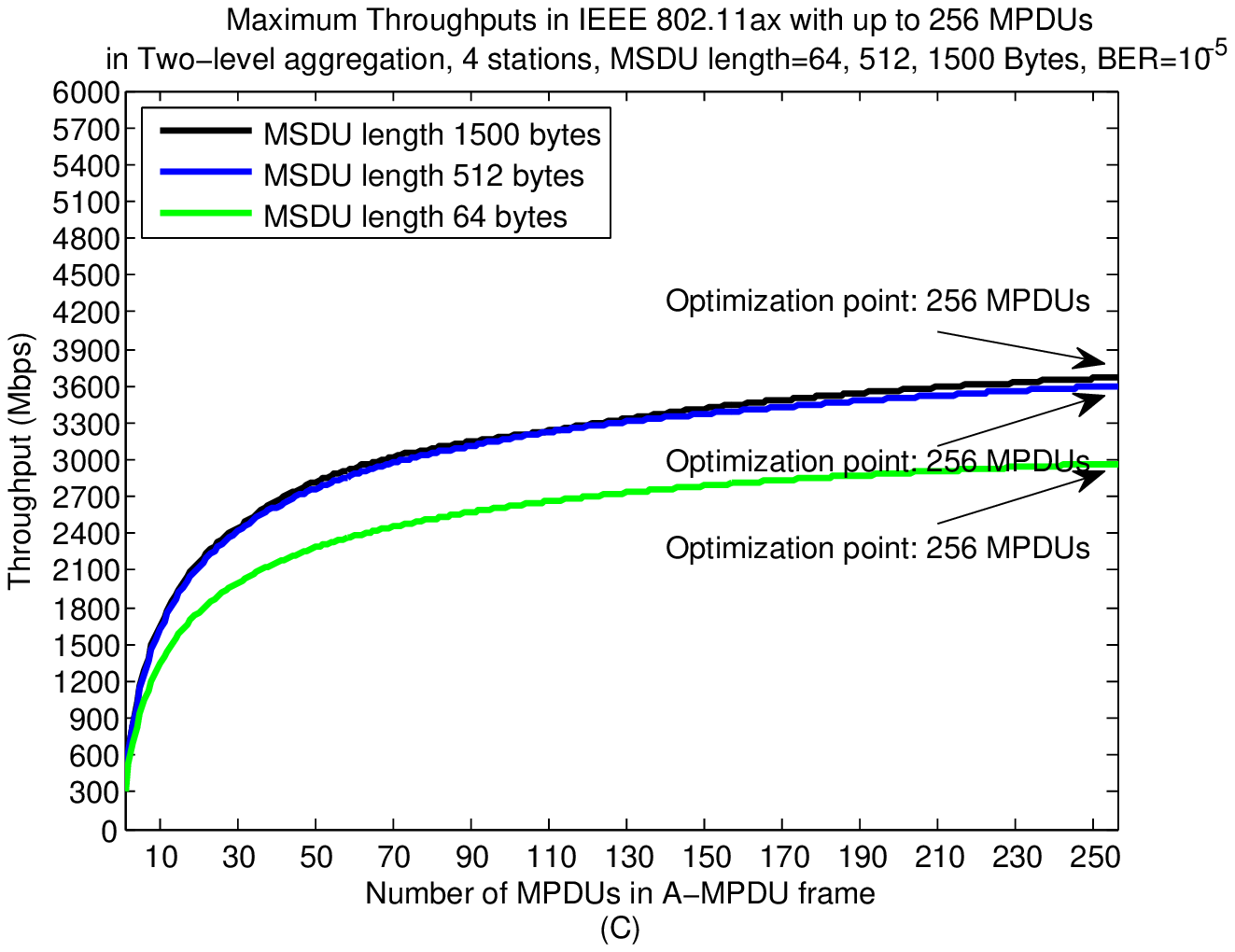}
\includegraphics{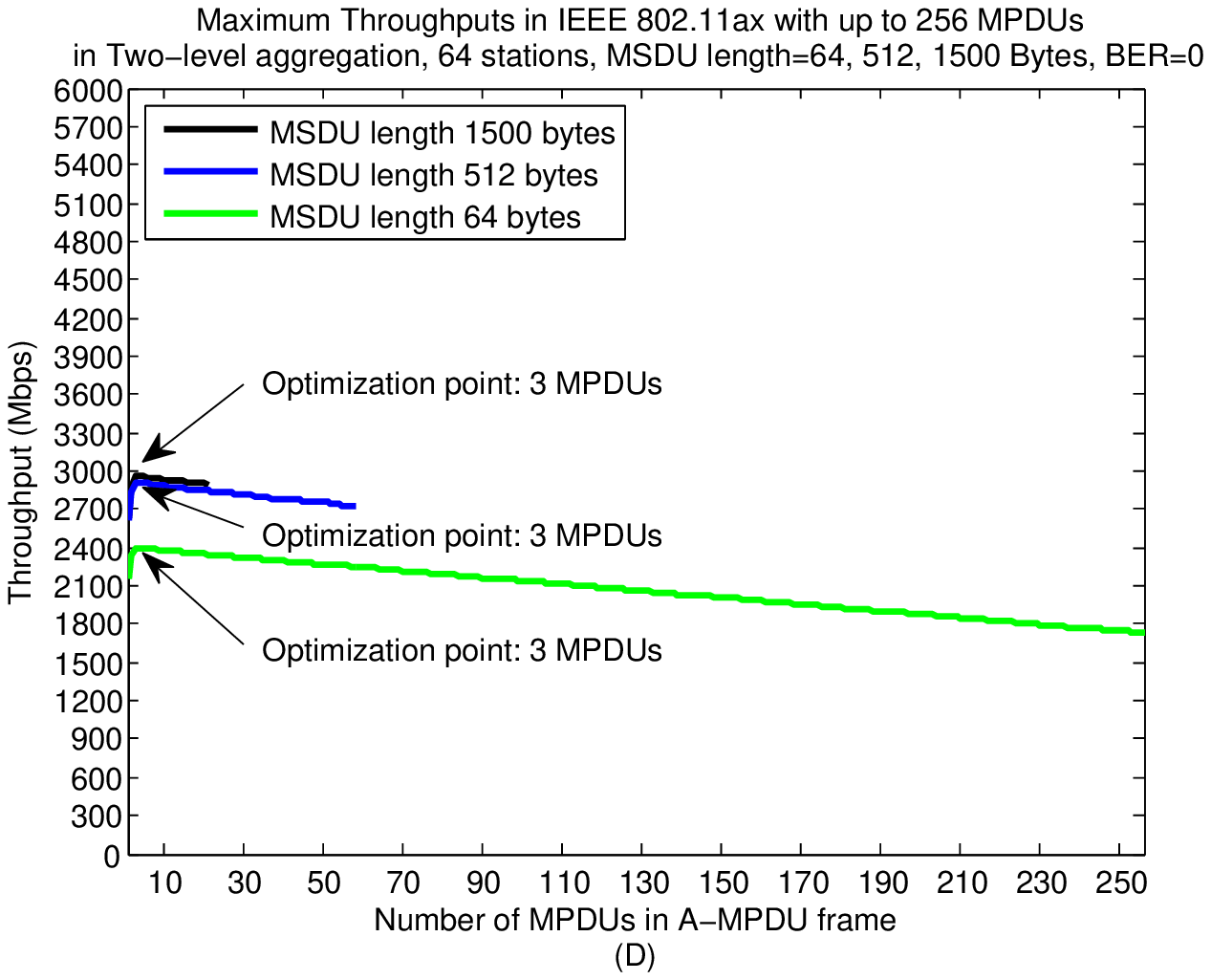}
\includegraphics{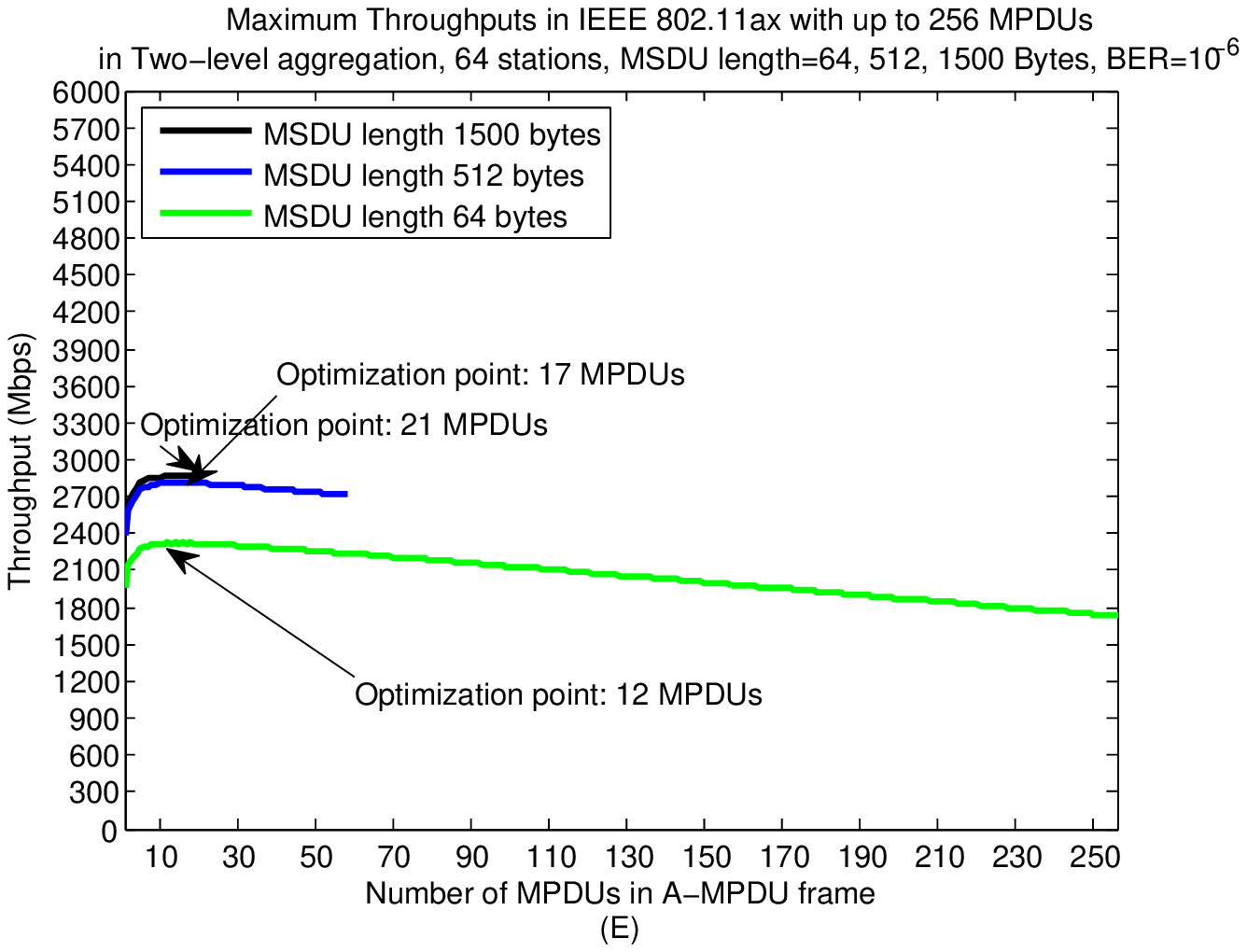}
\includegraphics{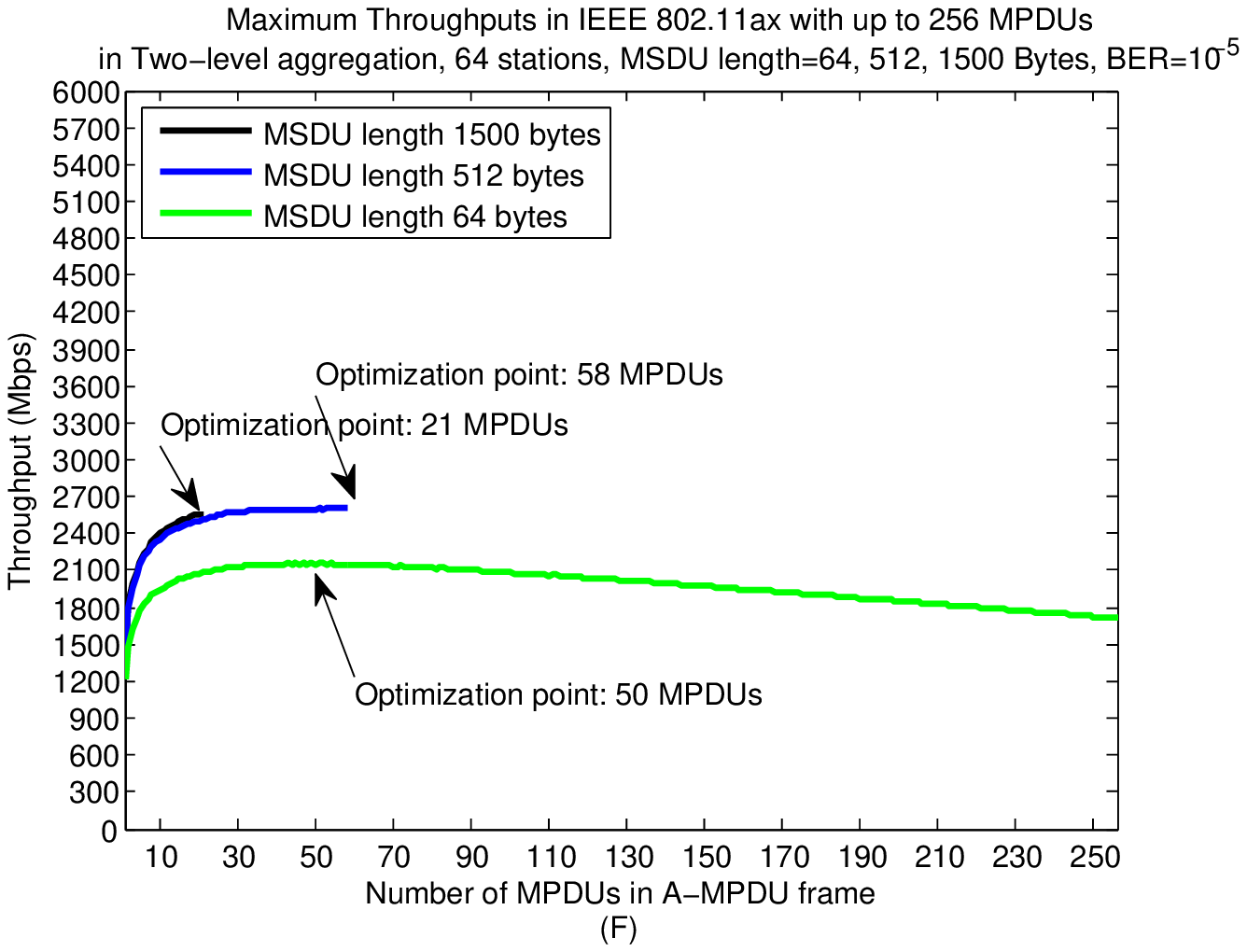}
\caption{The throughputs vs. the number of MPDUs in A-MPDU frames in IEEE 802.11ax Multi User for 4 stations in MCS11 and 64 stations in MCS9.}
\label{fig:res3}
\end{figure}

In Figure~\ref{fig:res3} we show the impact of the number
of MPDUs in A-MPDU frames on the received throughput.
In Figures~\ref{fig:res3}(A),~\ref{fig:res3}(B) and~\ref{fig:res3}(C)
results are
shown for $MU_{AX}(4)$ in MCS11, for BER$=$0, $10^{-6}$ and $10^{-5}$
respectively. Similar results are shown for $MU_{AX}(64)$ for MCS9
in Figures~\ref{fig:res3}(D),~\ref{fig:res3}(E) and~\ref{fig:res3}(F) 
respectively. We show results
for MSDUs of 64, 512 and 1500 bytes.
We again concentrate on BER$=$0 and BER$=$$10^{-5}$ only.

Considering $MU_{AX}(4)$ and BER$=$0, Figure~\ref{fig:res3}(A), there
is an optimal number of MPDUs of around 72 for all
sizes of the MSDUs. In BER$=$0 it
is efficient to transmit the largest possible MPDUs.
Around 72 MPDUs, all the MPDUs contain the
largest possible number of MSDUs and transmission
time is used efficiently. Above 72 MPDUs the limit
of $5.484 ms$ on the PPDU transmission time
and the MPDUs' overhead cause a smaller number
of MSDUs to be transmitted and the throughput decreases.

In the case of BER$=$$10^{-5}$, Figure~\ref{fig:res3}(C), the
optimal number of MPDUs is 256 since MPDUs are short 
(to increase the MPDUs' transmission success probability)
and there is
enough transmission time for 256 MPDUs in the A-MPDU frame;
every additional MPDU increases the throughput.

In $MU_{AX}(64)$, Figures~\ref{fig:res3}(D),~\ref{fig:res3}(E) 
and~\ref{fig:res3}(F), the PHY rates are smaller
and the limit on the PPDU transmission time does
not enable transmission of many MPDUs with MSDUs of 512 and 1500 bytes.
Up to 21 and 58 MPDUs of these sizes can be transmitted 
respectively for BER$=$$10^{-5}$, containing one MSDU. For BER$=$0 an
optimal number of 3 MPDUs yields the
maximum throughput for all MSDUs' sizes. A larger number
of MPDUs decreases the number of MSDUs transmitted
due to MPDUs' overhead and the throughput decreases.
In the case of BER$=$$10^{-5}$ the MPDUs are shorter,
and increasing the number of MPDUs increases the throughput
since more MSDUs are transmitted.
An exception is the case of 64 bytes MSDUs. In this
case it is possible to transmit 256 MPDUs and several
MSDUs can be transmitted in every MPDU. Increasing
the number of MPDUs in this case over 50 MPDUs decreases the number
of MSDUs transmitted with a decrease in the throughput.
In the Appendix we derive an approximation for the optimal
number of MPDUs in an A-MPDU as a function of the BER.

\section{Summary}

In this paper we compare between DL service
scheduling flavors to optimize throughputs
of 11ac and 11ax over the DL when considering
UDP like traffic and several DL service scheduling stations are transmitting
in the system. We also consider several transmission
flavors in 11ac and 11ax using MU-MIMO and OFDMA.
We look for upper bounds on the throughput received at
the MAC layer after
neutralizing any aspects
of the PHY layer as the relation between the BER
and the MCSs in use, the number
of Spatial Streams (SS) in use, channel correlation
when using MU-MIMO, the sounding protocol etc.

11ax outperforms 11ac by the order of several tenths
of percentage because it enables simultaneous transmissions
on both the DL and the UL while
11ac has this capability over the DL only, and
for 4 stations only.
Also, 11ax has larger PHY rates which also improve
its efficiency compared to 11ac.

In 11ax there is not one best DL service scheduling
transmission flavor.
$MU_{AX}(8)$ achieves good results in terms
of throughout, but $MU_{AX}(16)$ and $MU_{AX}(32)$
also achieve good throughput results,
but with significantly smaller access delay. 11ax achieves its
best throughputs in MCS11 in the case of up to 32 stations,
and in MCS9 in the case of 64 stations.

There is an optimal A-MPDU frame structure.
In $MU_{AX}(4)$ it is sufficient to transmit around
70 MPDUs and 256 MPDUs in an A-MPDU frame for BER$=$0
and BER$=$$10^{-5}$ respectively. For $MU_{AX}(64)$ these
numbers of MPDUs are smaller, around 3 for BER$=$0 and
21, 58 and 50
for MSDUs of 1500, 512 and 64 bytes respectively, 
due to smaller PHY rates.

Finally, using up to 256 MPDUs in an A-MPDU frame outperforms
the case of using up to 64 MPDUs in the cases where the
PHY rates are large and/or the channel is unreliable, i.e.
BER$=$$10^{-5}$.

\clearpage

%%%%%%%%%%%%%%%%%%%%%%%%%%%%%%%%%%%%%%%%%%%%%%%%%%%%%%%%%%

\bibliographystyle{abbrv}
\bibliography{main}

%%%%%%%%%%%%%%%%%%%%%%%%%%%%%%%%%%%%%%%%%%%%%%%%%%%%%%%%%%%%%%%%

\end{document}